\def\CII{\hbox{[C$\scriptstyle\rm II$]\,}}
\def\CI{\hbox{[C$\scriptstyle\rm I$]\,}}

\documentclass[fleqn,usenatbib]{mnras}

\usepackage{newtxtext,newtxmath}
\usepackage[T1]{fontenc}

\DeclareRobustCommand{\VAN}[3]{#2}
\let\VANthebibliography\thebibliography
\def\thebibliography{\DeclareRobustCommand{\VAN}[3]{##3}\VANthebibliography}

\usepackage{graphicx}	% Including figure files
\usepackage{amsmath}	% Advanced maths commands
\usepackage{multirow}
\def\be{\begin{equation}}
\def\ee{\end{equation}}

\def\gsim{\lower.5ex\hbox{\gtsima}} 
\def\lsim{\lower.5ex\hbox{\ltsima}} 
\def\gtsima{$\; \buildrel > \over \sim \;$} 
\def\ltsima{$\; \buildrel < \over \sim \;$} \def\gsim{\lower.5ex\hbox{\gtsima}} 
\def\lsim{\lower.5ex\hbox{\ltsima}} 
\def\simgt{\lower.5ex\hbox{\gtsima}} 
\def\simlt{\lower.5ex\hbox{\ltsima}}

\def\S*{$\Sigma_{\rm SFR}$}

\def\CII{\hbox{[C~$\scriptstyle\rm II $]~}}

\title[Power spectrum of  \CII halos]{The power spectrum of extended \CII halos around high redshift galaxies}

\author[M. Zhang et al.]{
Meng Zhang,$^{1,2,3}$,
Andrea Ferrara$^{3}$\thanks{E-mail: andrea.ferrara@sns.it}, Bin Yue$^{1, 4}$
\\
$^{1}$National Astronomical Observatories, Chinese Academy of Sciences, 20A, Datun Road, Chaoyang District, Beijing, 100101, China\\
$^{2}$School of Astronomy and Space Science, University of Chinese Academy of Sciences, No.1 Yanqihu East Rd, Huairou District, Beijing 101408, China \\
$^{3}$Scuola Normale Superiore, Piazza dei Cavalieri 7, 56126 Pisa, Italy\\
$^{4}$Key Laboratory of Radio Astronomy and Technology, Chinese Academy of Sciences, 20A Datun
Road, Chaoyang District, Beijing 100101, China
}

\date{Accepted XXX. Received YYY; in original form ZZZ}

\pubyear{2023}

\begin{document}
\label{firstpage}
\pagerange{\pageref{firstpage}--\pageref{lastpage}}
\maketitle

% Abstract of the paper
\begin{abstract}
ALMA observations have detected extended ($\simeq 10$ kpc) \CII halos around high-redshift ($z\simgt 5$) star-forming galaxies. If such extended structures are common, they may have an impact on the line intensity mapping (LIM) signal. We compute the LIM power spectrum including both the central galaxy and the \CII halo, and study the detectability of such signal in an ALMA LIM survey. We model the central galaxy and the \CII halo brightness with a $\rm S\acute{e}rsic$+exponential profile. The model has two free parameters: the effective radius ratio $f_{R_{\rm e}}$, and the central surface brightness ratio, $f_{\Sigma}$, between the two components. \CII halos can significantly boost the LIM power spectrum signal. For example, for relatively compact \CII halos ($f_\Sigma=0.4$, $f_{R_{\rm e}}=2.0$), the signal is boosted by $\simeq 20$ times; for more extended and diffuse halos ($f_\Sigma=0.1, f_{R_{\rm e}}=6.0$), the signal is boosted by $\simeq 100$ times.
For the ALMA ASPECS survey (resolution $\theta_{\rm beam} = 1.13''$, survey area $\Omega_{\rm survey}=2.9\,\rm arcmin^{2}$) the \CII  power spectrum is detectable only if the deL14d \CII - SFR relation holds. However, with an optimized survey ($\theta_{\rm beam} = 0.232''$, $\Omega_{\rm survey}=2.0\,\rm deg^{2}$), the power spectrum is detectable for all the \CII - SFR relations considered in this paper. Such a survey can constrain $f_\Sigma$ ($f_{R_{\rm e}}$) with a relative uncertainty of $\sim15\%$ ($\sim 10\%$). A successful LIM experiment will provide unique constraints on the nature, origin, and frequency of extended \CII halos, and the \CII - SFR relation at early times.

\end{abstract}

% Select between one and six entries from the list of approved keywords.
% Don't make up new ones.

%%%%%%%%%%%%%%%%%%%%%%%%%%%%%%%%%%%%%%%%%%%%%%%%%%

%%%%%%%%%%%%%%%%% BODY OF PAPER %%%%%%%%%%%%%%%%%%

\section{Introduction}
Cosmic reionization, as the last major phase transition of the Universe, is a direct consequence of the formation of the first luminous objects. Determining the properties of galaxies in the Epoch of Reionization (EoR) is crucial to understand structure formation and evolution \citep[][]{Dayal+2018}. However, our progress is hampered by the limited number of EoR sources that can be directly accessed via targeted observations. Such limitation is particularly severe for faint galaxies, which are thought to be the primary sources of ionizing photons \citep[][]{Robertson+2015, Mitra15, Castellano+2016} but are far below the detection limits of the current surveys \citep[][]{Salvaterra+2013}.

Line intensity mapping (LIM) is emerging as an efficient tool to detect these faint galaxies. LIM measures the integrated emission of spectral lines from all galaxies and the intergalactic medium (IGM) \citep[see a review by][]{Kovetz+2017}. By measuring fluctuations in the line emission from early galaxies, one could expect to obtain valuable physical insights about the properties of these sources and their role in reionization. 
Finally, LIM can also be used to put stringent constraints on cosmological models \citep[][]{Kovetz+2019, Schaan+2021, Karkare+2022, Bernal+2022}. For example, \citet{Gong+2020} proposed to use the multipole moments of the redshift-space LIM power spectrum to constrain the cosmological and astrophysical parameters.

There are several spectral lines of interest for intensity mapping including HI 21 cm line \citep[][]{Chang+2010, Salvaterra+2013}, CO rotational lines \citep[][]{Breysse+2014, Mashian+2015, Li+2016}, bright optical emission lines such as $\rm Ly \alpha$ and $\rm H \alpha$ \citep[][]{Salvaterra+2013, Pullen+2014, Comaschi+2016, Gong+2017, Silva+2018} and the far-infrared (FIR) fine-structure lines \citep[][]{Gong+2012, Uzgil+2014, Silva+2015, Yue+2015, Serra+2016, Yue+2019}.

Among these, the \CII emission line with wavelength 157.7 $\mu$m, corresponding to the $^{2}P_{3/2} \rightarrow ^{2}P_{1/2}$ forbidden transition of singly ionized carbon, is the brightest one in the FIR  \citep[][]{Stacey+1991}, and a dominant coolant of the neutral ISM \citep[][]{Stacey+1991, Wolfire+2003} and dense photo-dissociation regions \citep[PDR,][]{Hollenbach+1999}. Therefore, the \CII line can be used to probe the properties of the ISM at high redshift \citep[][]{Capak+2015, Pentericci+2016, Knudsen+2016, Carniani+2017, Bakx+2020, Matthee+2020}.

Notably, a tight relation between \CII luminosity and star formation rate (SFR) is found from both observations \citep[][]{DeLooze+2014, Herrera-Camus+2015, Schaerer+2020} and simulations \citep{Vallini+2015, Olsen+2017, Leung+2020}. 
Therefore, the \CII emission lines can be used to trace the star formation across cosmic time, albeit at high redshift the scatter around the local relation increases \citep[][]{Carniani+2018}. Conveniently, at early epochs, the line is shifted into the (sub-)mm wavelength range, which is accessible to ground-based telescopes such as the Atacama Large Millimeter/sub-millimeter Array (ALMA).

With the advent of ALMA, a large number of galaxies with \CII emission line at $z>4$ have been detected, boosting the studies of the obscured star formation at high redshift \citep[][]{Hodge+2020}. Among these, a stacking analysis of ALMA observed galaxies \citep[][]{Fujimoto+2019} discovered extended, 10 kpc scale \CII halos around high redshift galaxies, whose size is $\simeq 5$ times larger than the UV size of the central galaxy. Similar results have been reported in the following studies both in subsequent stacking analysis \citep{Ginolfi+2020, Fudamoto+2022} and individual galaxies \citep{Fujimoto+2020, Herrera-Camus+2021, Akins+2022, Lambert+2022, Fudamoto+2023}. 

The existence of extended \CII halos around normal star-forming galaxies opens new perspectives for early metal enrichment. However, the physical origin of \CII halos is still not clear. Possible scenarios include satellite galaxies, extended PDR or HII regions, cold streams, and outflows. These are discussed in \citet[][]{Fujimoto+2019}. 
Indeed, the supernova-driven cooling outflow model explored by \citet[][]{Pizzati+2020, Pizzati23} successfully produces the extended \CII halo, their model also indicates that outflows are widespread phenomena in high-z galaxies, but the extended \CII halos for low-mass galaxies are likely too faint to be detected with present levels of sensitivity. The outflow scenario is also supported by \citet[][]{Ginolfi+2020}, who found outflow signatures in the stacking analysis of \CII emission
detected by ALMA in 50 main sequence star-forming galaxies at $4<z<6$. \citet[][]{Fujimoto+2020} also suggested that the star-formation-driven outflow is the most likely origin of the \CII halos. Finally, \cite{Herrera-Camus+2021} found evidence of outflowing gas, which may be responsible for the production of extended \CII halos. On the other hand, \citet{Heintz+2023} use \CII emission as a proxy to infer the metal mass in the
interstellar medium (ISM) of galaxies, and find that the majority of metals produced at $z \gtrsim 5$ are confined to the ISM, disfavour efficient outflow
processes at these redshifts, they claim extended \CII halos trace the extended neutral gas
reservoirs of high-z galaxies. In any case, if \CII halos are common in high redshift normal star-forming galaxies, they could leave a distinct imprint in LIM experiments.

In this paper, we aim to investigate the effects of extended \CII halos on the LIM signal and their detectability. We first construct the \CII halo model and compute the intensity mapping power spectrum of high redshift galaxy systems when both the central galaxy and extended \CII halo are considered\footnote{Throughout the paper, we assume $\Lambda$CDM model with  \citet[][]{Planck2016} cosmological parameters: $\Omega_{\rm m} = 0.308$, $\Omega_{\rm \Lambda}=1-\Omega_{\rm m} = 0.692$, $\Omega_{\rm b} = 0.048$, $h=0.678$, $\sigma_{\rm 8} = 0.815$, $n_{\rm s} = 0.968$.}. 
Then we analyze the detectability of the signal in the ALMA intensity mapping survey and a proposed optimized survey. The paper is organized as follows: We outline our method in Section \ref{sec:method}. Results, including the power spectrum signal and its signal-to-noise ratio (S/N) estimation, are presented in Section \ref{sec:results}. We finally summarize the results and give a discussion in Section \ref{sec:conclusion}.

\section{Method}\label{sec:method}

In the following, we describe our model to derive the LIM signal when both central galaxies and \CII halos are considered. We start by constructing the extended \CII halo model, and the relations between central galaxy \CII luminosity, extended \CII halo luminosity, and dark matter halo mass. We then compute the intensity mapping power spectrum, including one-halo and two-halo terms, and the shot noise term. Finally, we introduce the method to estimate the signal-to-noise ratio of the \CII power spectrum, given the LIM survey parameters.

\subsection{\CII halo model}

To compute the clustering term of the power spectrum, the first step is to model the \CII radial surface brightness of the central galaxy and the extended halo. For this, we use a combined $\rm S\acute{e}rsic$+exponential model, where the central galaxy is described by the $\rm S\acute{e}rsic$ model \citep[][]{Sersic+1963, Sersic+1968} while the extended \CII halo is described by the exponential function \citep{Fujimoto+2019, Akins+2022}.

The \CII radial surface brightness of the central galaxy writes
\begin{equation}
    \Sigma_{\rm CII, g} (R)=C_{\rm g} \exp \left[-b_n \left(\frac{R}{R_{\rm e,g}} \right)^{1/n}\right] ,
    \label{eq:Sersic}
\end{equation}
where $C_{\rm g}$ is the central surface brightness, $R$ is the projected distance to the source centre in the plane-of-sky, $n$ is the $\rm S\acute{e}rsic$ index, and $R_{\rm e,g}$ is the effective radius containing half of the integrated brightness. The term $b_n$ is a function of $n$. It is obtained by solving the equation $\Gamma(2n)=2\gamma(2n, b_n)$, where $\Gamma(2n)$ is the Gamma function, $\gamma(2n,b_n) = \int^{ b_n}_{0} t^{2n-1} e^{ -t} dt$ is the lower incomplete gamma function.  

In the stacking analysis by \citet{Fujimoto+2019}, the central galaxies are well fitted by $\rm S\acute{e}rsic$ model with $n=1.2$ and $R_{\rm e,g}=1.1 \,\rm kpc$. This is consistent with the rest-frame optical and UV sizes measured by \citet{Shibuya+2015}, who obtain a nearly constant value of $R_{\rm e,g}/R_{\rm vir}= 1.0\%-3.5\% $ for $z = 0-8$, where $R_{\rm vir}$ is the dark matter halo virial radius. Following these results, we adopt $n =1.2 $ and fix $R_{\rm e,g} = 0.03 R_{\rm vir}$. 

For the extended \CII halo, the surface brightness has the exponential form\footnote{Note that the $\rm S\acute{e}rsic$ model Eq. (\ref{eq:Sersic}) is reduced to exponential function when $n=1$.
}
\begin{equation}
    \Sigma_{\rm CII, h} (R) = 
    C_{\rm h} \exp \left[-b_1 \frac{R}{R_{\rm e, h}}   \right],
\end{equation}
where $b_1$ is the $b_n$ for $n=1$, $C_{\rm h}$ is the surface brightness at the center and the $R_{\rm e,h}$ is the effective radius of the \CII halo. We further assume $C_{\rm h}=f_\Sigma C_{\rm g}$ and  $R_{\rm e,h}=f_{R_{\rm e}} R_{\rm e,g}$, where the ratios $f_\Sigma$ and $f_{R_{\rm e}}$ are two free parameters in our work.  

With the above surface brightness profiles, the \CII luminosity for the central galaxy 
\begin{align}
     L_{\rm CII,g} &=   \int 2\pi  R \Sigma_{\rm CII, g}(R) dR \nonumber \\
     &= 2\pi C_{\rm g} \frac{n \Gamma(2n)}{(b_n)^{2n}} R_{\rm e, g}^{2},
     \label{eq:central luminosity}
\end{align}
and for the extended \CII halo 
\begin{align}
    L_{\rm CII,h} &=  \int 2\pi R \Sigma_{\rm CII, h}(R) dR \nonumber \\
    &= 2\pi C_{\rm h} \frac{\Gamma(2)}{(b_1)^{2}} R_{\rm e,h}^{2}.
\end{align}

Since restricted by the sensitivity of the telescope in previous observations, the observed \CII luminosity in deriving the \CII - SFR relation is likely dominated by central galaxies. Moreover, as noted by \cite{Fujimoto+2019}, to fully capture the extended \CII halos, additional mechanisms not involved in previous simulations are required. 
Therefore, we consider the previously predicted \CII - SFR relations (both by observations and by simulations ) do not account for (or at least significantly underestimate) the extended \CII halo contribution. 
For clarification, we shall use $L_{\rm CII, SFR}-{\rm SFR}$ to denote the relations in the following of this work, where $L_{\rm CII, SFR}$ is the \CII luminosity derived from the previously \CII - SFR relations, see next subsection.
We find the normalization by assigning $L_{\rm CII, g} \approx L_{\rm CII, SFR}$.
Then
\begin{equation}
    C_{\rm g} = \frac{ (b_n)^{2n}}{2 \pi n \Gamma(2 n) R_{\rm e, g}^{2} } L_{\rm CII,SFR}.
\end{equation}

Then the \CII halo luminosity is also derived from the $L_{\rm CII,SFR}$ via
\begin{equation}
    L_{\rm CII,h} = f_{\Sigma} f_{R_{\rm e}}^{2} \frac{(b_n)^{2n}}{(b_1)^{2}} \frac{\Gamma(2)}{n\Gamma(2n)} L_{\rm CII, SFR},  
    \label{eq:halo luminosity}
\end{equation}
and when $n=1.2$ it is reduced to 
\begin{equation}
    L_{\rm CII,h}  \simeq 2.24 f_{\Sigma} f_{R_{\rm e}}^{2} L_{\rm CII, SFR}.
    \label{eq:halo luminosity numerical}
\end{equation}

\begin{figure*}
	\includegraphics[width=2\columnwidth]{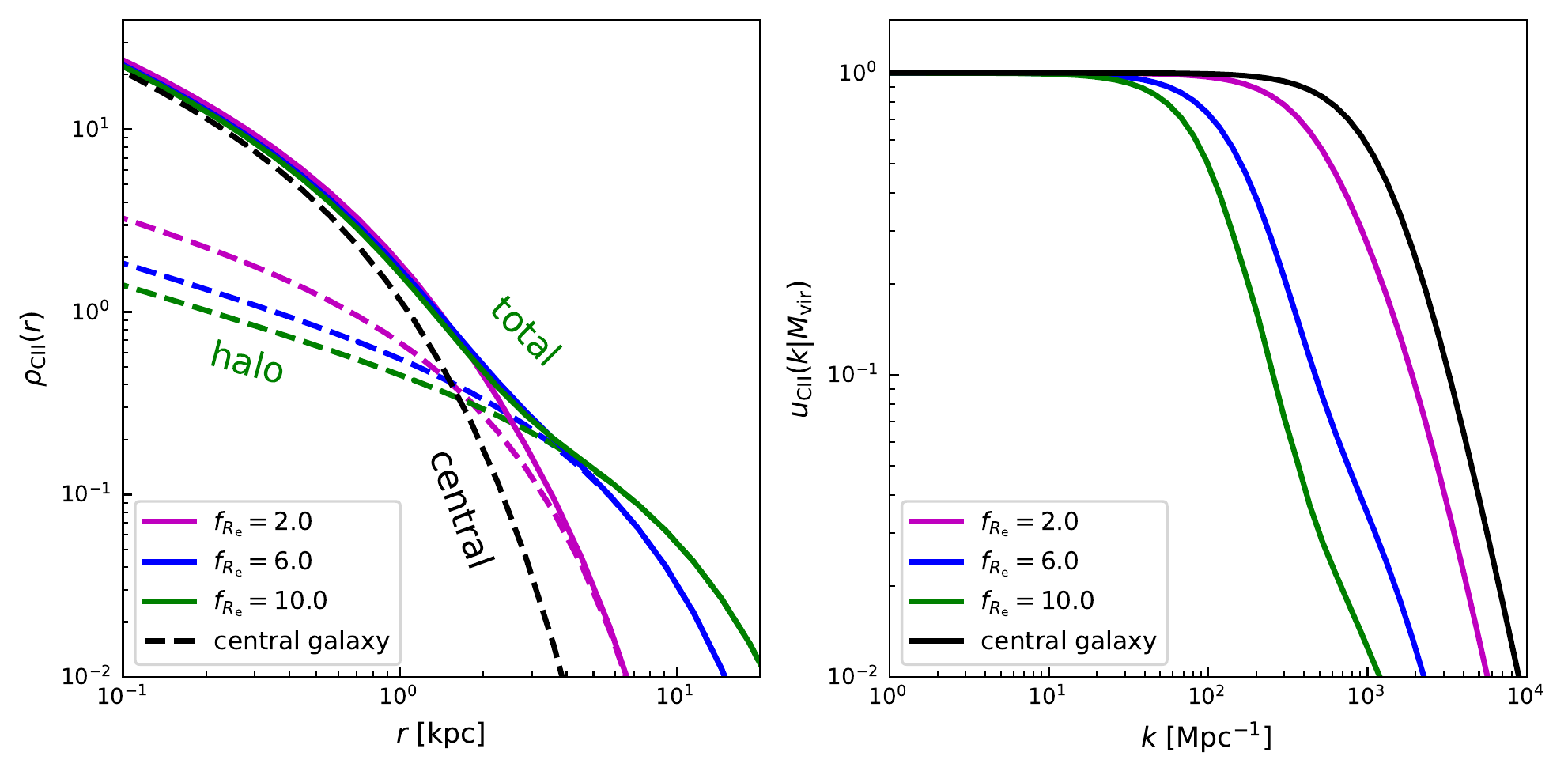}
    \caption{ \textit{Left}: The radial profile of a \CII halo in dark matter halo with mass $M_{\rm vir}=10^{12}\,\rm M_{\odot}$. We fix \CII halo parameter $f_{\Sigma}=0.4$ and vary the effective radius ratios $f_{R_{\rm e}}$. The black dashed line is the profile of the central galaxy. The green, blue, and magenta dashed lines are the profiles of the \CII halo part with $f_{R_{\rm e}} = [2.0, 6.0, 10.0]$ respectively. The solid lines represent the sum of the central galaxy and \CII halo components. \textit{Right}: The corresponding normalized Fourier transform of the \CII halo profile with different effective radius ratios $f_{R_{\rm e}}$.}  
    \label{fig:density_profile}
\end{figure*}

We further assume that the profile is spherically symmetric, and perform a deprojection of the above surface density to obtain the 3D brightness profiles for the two components, $\rho_{\rm CII,g}(r)$ and $\rho_{\rm CII, h}(r)$. This is given in Appendix~\ref{sec:deprojection}. The total 3D brightness profile is the sum of the two components
\begin{equation}
    \rho_{\rm CII}(r) = \rho_{\rm CII,g}(r) + \rho_{\rm CII, h}(r),
\end{equation}
 and we truncate it at $R_{\rm vir}$. Then, the normalized Fourier transform of the profile in a dark matter halo of mass $M_{\rm vir}$, 
\begin{equation}
u_{\rm CII}(k|M_{\rm vir})=\frac{\int^{R_{\rm vir}}_{0} dr \,4 \pi r^{2} \frac{\sin{kr}}{kr} \rho_{\rm CII}(r)} {\int^{R_{\rm vir}}_{0} dr\, 4\pi r^{2} \rho_{\rm CII}(r)},
\end{equation}
which is used to compute the one-halo and two-halo terms of the power spectrum. The luminosity density profiles of the central galaxy and \CII halo are shown in the left panel of Fig. \ref{fig:density_profile}. We take $M_{\rm vir}=10^{12}~M_\odot$ at $z=6$, $f_\Sigma=0.4$ and $f_{R_{\rm e}}=2.0, 6.0, 10.0$. In the right panel of Fig. \ref{fig:density_profile} we show the Fourier transform of the profiles for different $f_{R_{\rm e}}$.

\subsection{The \CII- dark matter halo mass relation}\label{sec:CII DM relation}
To derive the $L_{\rm CII, SFR}-M_{\rm vir}$ relation, we first use the abundance matching method to get $\rm SFR-M_{\rm vir}$ relation. Following \citet[][]{Yue+2015}, we start from the observed (dust-attenuated) galaxy UV luminosity function (LF), which can be described by a Schechter function \citep[][]{Schechter+1976}
\begin{equation}
\begin{aligned}
    \frac{dn}{dM_{\rm UV}^{\rm obs}} &= 0.4 \ln(10) \Phi^{*} 10^{-0.4(M_{\rm UV}^{\rm obs}-M_{\rm UV}^{*})(\alpha +1)}\\
    & \times \exp{[-10^{-0.4(M_{\rm UV}^{\rm obs}-M_{\rm UV}^{*})}]},
\end{aligned}
\end{equation}
where the $M_{\rm UV}^{\rm obs}$ is the dust-attenuated absolute UV magnitude, and the redshift-dependent parameters are taken from \citet{Bouwens+2015},
\begin{equation}
\begin{aligned}
    M_{\rm UV}^{*} &= -20.95 + 0.01 (z-6),\\
    \Phi^{*} &= 0.47 \times 10^{-3-0.27 (z-6)},\\
    \alpha &= -1.87-0.10 (z-6). 
\end{aligned}
\end{equation}

Considering dust attenuation, the intrinsic absolute UV magnitude becomes 
\begin{equation}
    M_{\rm UV} = M_{\rm UV}^{\rm obs} - A_{\rm UV},
\end{equation}
where the UV dust attenuation is parameterized as 
\begin{equation}
    A_{\rm UV} = C_{1} + C_{0} \beta\  (A_{\rm UV}\geq 0);
\end{equation}
we set the coefficients $C_{0}=2.10$ and $C_{1} = 4.85$, following \citet{Koprowski+2018}. Finally, the UV spectral slope $\beta$ is fitted by \citet{Bouwens+2015}
\begin{equation}
    \beta = \beta_{\rm 0} + \frac{d\beta}{d M_{\rm 0}} (M_{\rm UV}^{\rm obs}-M_{\rm 0}),
\end{equation}
with a redshift dependence given by 
\begin{equation}
\begin{aligned}
    \beta_{0} = -1.97-0.06(z-6), \\
    \frac{d \beta}{dM_{\rm UV}} = -0.18-0.03 (z-6).
\end{aligned}
\end{equation}

The intrinsic UV LF is related to the observed UV LF by 
\begin{equation}
    \frac{dn}{dM_{\rm UV}} = \frac{dn}{dM_{\rm UV}^{\rm obs}} \frac{dM_{\rm UV}^{\rm obs}}{dM_{\rm UV}}. 
\end{equation}

By assuming that the most massive galaxies occupy the most massive dark matter halos, we use the abundance matching method
\begin{equation}
    \int_{M_{\rm UV}} \frac{dn}{dM_{\rm UV}} dM_{\rm UV} = \int_{M_{\rm vir}} \frac{dn}{dM_{\rm vir}} dM_{\rm vir}, 
\end{equation}
to derive the intrinsic absolute UV magnitude and dark matter halo mass relation. To get the ${\rm SFR}-M_{\rm vir}$ relation, we assume that the intrinsic UV luminosity scales with the SFR \citep[][]{Kennicutt+1998}
\begin{equation}
    {\rm SFR} = K_{\rm UV} L_{\rm UV}, 
\end{equation}
where $K_{\rm UV} = 0.7 \times 10^{-28}\, \rm M_{\odot}\ yr^{-1}/(erg\ s^{-1}\ Hz^{-1})$ \citep{Bruzual+2003} for a \citet{Chabrier+2003} stellar initial mass function (IMF), metallicity in the range $0.005-0.4\ Z_{\odot}$, and stellar age $>100\,\rm Myr$; we ignore the scatter on $K_{\rm UV}$ as discussed by \citet{Yue+2019}. 

The final step to get the $L_{\rm CII, SFR}-M_{\rm vir}$ relation involves the $L_{\rm CII, SFR}-\rm SFR$ relation. This can be parameterized as 
\begin{equation}
\begin{aligned}
\log L_{\rm CII,SFR} &= \log \bar{L}_{\rm CII, SFR} \pm \sigma_{\rm L}\\
&= \log A + \gamma \log \rm{SFR} \pm \sigma_L,
\end{aligned}
\end{equation}
where SFR is in units of $\rm M_{\rm \odot}\ {\rm yr}^{-1}$, and $L_{\rm CII,SFR}$ in units of $\rm L_{\odot}$. We adopt six different $L_{\rm CII, SFR}-\rm SFR$ relations to compute the power spectrum in our model. These relations are described in the following.

By combining a semi-analytical model of galaxy formation with the photo-ionisation code CLOUDY \citep[][]{Ferland+2013, Ferland+2017} to compute the \CII luminosity for a large number of galaxies at $z \geq 4$, \citet[][hereafter L18]{Lagache+18} reproduced the $L_{\rm CII, SFR}-\rm SFR$ relation observed in high-$z$ star-forming galaxies. They found a mild evolution of the $L_{\rm CII,SFR}-\rm SFR$ relation with redshift from $z=4$ to $z=7.6$, 
\begin{equation}
    \log L_{\rm CII,SFR} = (7.1-0.07z)+(1.4-0.07z)\log(\rm SFR),
\rm \end{equation}
with a scatter of about 0.5 dex.

\citet{DeLooze+2014} analysed the $L_{\rm CII, SFR}-\rm SFR$ relation for low-metallicity dwarf galaxies based on Herschel observations \citep{Madden+2013} and found a local relation with $\log A = 7.16$, $\gamma = 1.25$ and 0.5 dex scatter (hereafter deL14d), i.e.
\begin{equation}
    \log L_{\rm CII, SFR} = 7.16 + 1.25 \log \rm SFR,
\end{equation}

\citet{DeLooze+2014} also obtained $\log A = 7.22$, $\gamma = 0.85$  and 0.3 dex scatter (hereafter deL14$z$) for high-$z$ samples with $z = 0.59-6.60$. 
\begin{equation}
    \log L_{\rm CII, SFR} = 7.22 + 0.85 \log \rm SFR
\end{equation}

\citet{Olsen+2017} obtained $\log A = 6.69 $ and $\gamma = 0.58$, with a scatter of $\sim 0.15$ dex (here after O17) by combining cosmological zoom simulations of galaxies with S\'{I}GAME \citep[][]{Olsen+2015} to model \CII emissions from 30 main-sequence galaxies at $z \sim 6$,
\begin{equation}
    \log L_{\rm CII,SFR} = (6.69 \pm 0.10)+(0.58 \pm 0.11) \log \rm SFR. 
\end{equation}
This is consistent with \citet{Leung+2020} who predict $\log A=6.82$ and $\gamma=0.66$, based on cosmological hydrodynamics simulations using the SIMBA suite plus radiative transfer calculations via an updated version of S\'{I}GAME.

\citet{Vallini+2015} combined high-resolution, radiative transfer cosmological simulations with a subgrid multiphase model of the ISM to model the \CII emissions from diffuse neutral gas and PDRs. By considering a physically-motivated metallicity, they found $L_{\rm CII, SFR}-\rm SFR$ relation both depends on the SFR and metallicity and can be fitted by 
\begin{equation}
    \begin{aligned}
    \log {L}_{\rm CII,SFR} &= 7.0+1.2 \log(\rm SFR) + 0.021  \log(Z)\\
    &+ 0.012  \log(\rm SFR) \log(Z) -0.74  \log^{2} (Z).
    \end{aligned}
\end{equation}
For $Z = 0.2\, \rm Z_{\odot}$, we obtain $\log A = 6.62$ and $\gamma = 1.19$ (here after V15).  

\citet{Schaerer+2020}, using 118 galaxies at $4.4 < z < 5.9$ from the ALMA ALPINE Large Program \citep{LeFevre+2020}, obtain 
\begin{equation}
    \log L_{\rm CII,SFR} = \left(6.61 \pm 0.2 \right) + \left(1.17 \pm 0.12 \right) \log \rm SFR,
\end{equation}
with a scatter of $\sim 0.28$ dex (here after S20).

\begin{figure}
	\includegraphics[width=\columnwidth]{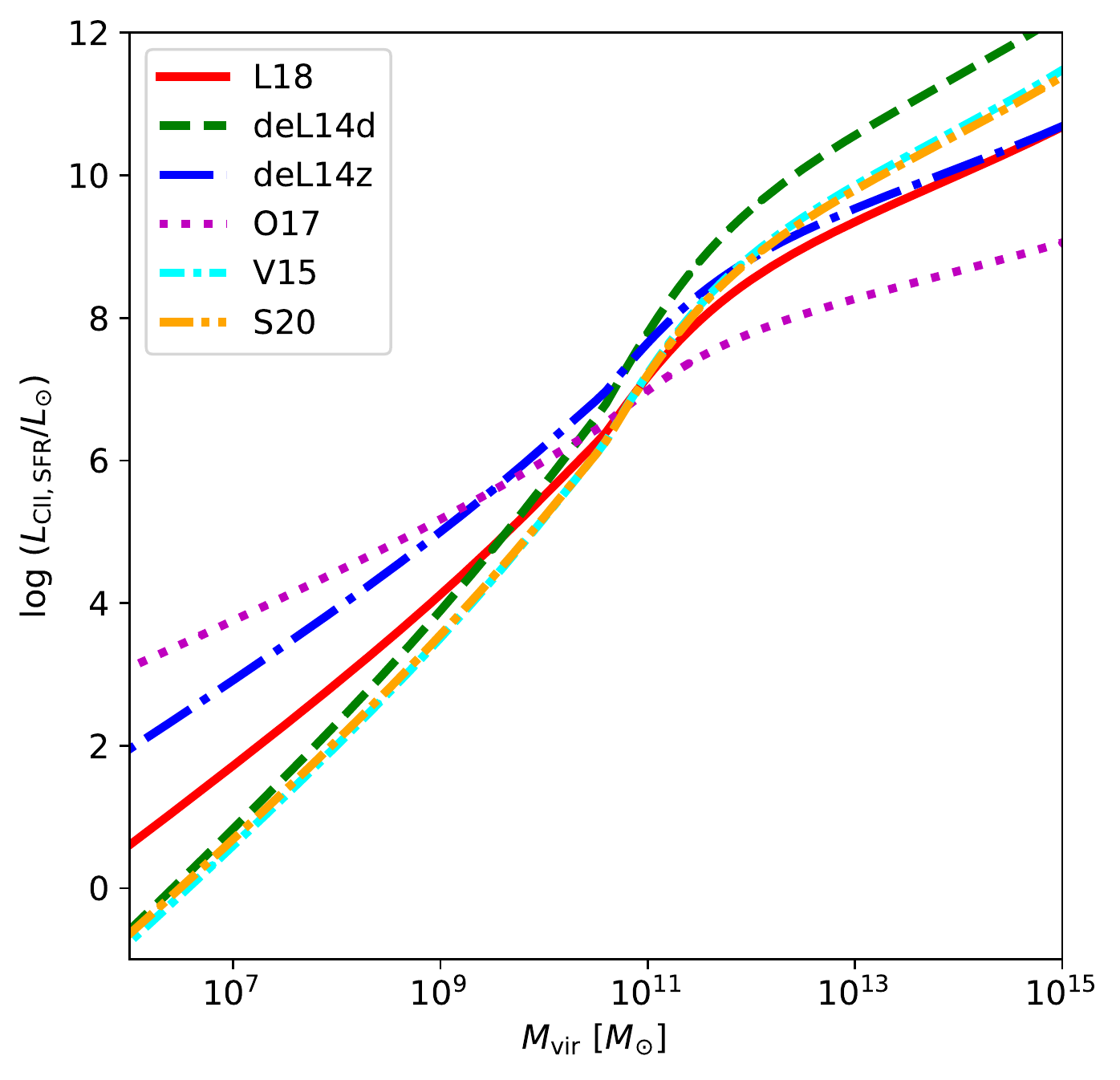}
    \caption{The $L_{\rm CII, SFR}-M_{\rm vir}$ relation obtained from Abundance Matching method for six $L_{\rm CII, SFR}-\rm SFR$ relations.}
    \label{fig:LCII_Mh}
\end{figure}

The parameters of the six $L_{\rm CII, SFR} - {\rm SFR}$ relations adopted are summarized in Tab.~\ref{tab:CII SFR relation}. Since the observed $\sigma_{\rm L}$  involves both the observational uncertainties and the intrinsic scatters, and the simulated $\sigma_{\rm L}$ might be limited by the method of modelling the input physical properties scattering, for simplicity in this work we ignore the $\sigma_{\rm L}$ and simply use $\log L_{\rm CII, SFR}=\log A+\gamma \log {\rm SFR}$ in the following. 
We plot the corresponding $L_{\rm CII,SFR}-M_{\rm vir}$ relations in Fig.~\ref{fig:LCII_Mh}.

\begin{table}
	\centering
	\caption{Summary of $L_{\rm CII, SFR}-\rm SFR$ relation parameters adopted in our model.}
	\label{tab:CII SFR relation}
	\begin{tabular}{*{6}{c}} % four columns, alignment for each
		\hline
		model & $\log\ A$ & $\gamma$ & $\sigma_{\rm L}$ & Redshift & References\\
		\hline
		L18 & 6.68 & 0.98 & 0.60 & ~6 & \citet[][]{Lagache+18}\\
		
		deL14d & 7.16 & 1.25 & 0.5 & Local & \citet[][]{DeLooze+2014}\\
		
		deL14z & 7.22 & 0.85 & 0.3 & 6.6 & \citet[][]{DeLooze+2014}\\
		
		O17 & 6.69 & 0.58 & 0.15 & 6 & \citet[][]{Olsen+2017}\\
		
		V15 & 6.62 & 1.19 & 0.4 & ~6.6 & \citet[][]{Vallini+2015}\\

		S20 & 6.61 & 1.17 & 0.28  & 4.4-5.9 & \citet[][]{Schaerer+2020}\\
 		
		\hline
		
	\end{tabular}
\end{table}

\subsection{Modeling the power spectrum}
Since galaxies are biased and discrete tracers of the dark matter density fluctuations, basically the LIM power spectrum comprises clustering and shot noise components \citep{Kovetz+2017}. The clustering term describes the large-scale clustering nature of galaxies and the shot noise term originates from the Poisson fluctuations of galaxy numbers. In the halo model framework \citep{Cooray+2002}, the clustering term can be split into one-halo term and two-halo term, which arise from the correlation within halos and between halos, respectively \citep{Moradinezhad-Dizgah2022}. The LIM power spectrum writes: 
\begin{equation}
\begin{aligned}
    P_{\rm CII} (k, z) &= P_{\rm CII}^{\rm CL}(k, z) + P_{\rm CII}^{\rm SN}(z)\\
    &= P_{\rm CII}^{\rm 1h}(k, z) + P_{\rm CII}^{\rm 2h}(k,z) + P_{\rm CII}^{\rm SN}(z),
    \label{eq:P_CII},
\end{aligned}
\end{equation}
where the shot noise $P_{\rm CII}^{\rm SN}(z)$ is generally independent of $k$. The one-halo term $P_{\rm CII}^{\rm 1h} (k, z)$ and the two-halo term $P_{\rm CII}^{\rm 2h} (k, z)$ depend on the large-scale structure, the normalized Fourier transform of the \CII halo profile and the $L_{\rm CII, SFR}-M_{\rm vir}$ relation.

Analogous to the halo model \citep{Cooray+2002} which assumes the correlation is tightly related to the (density or light) distribution profile within dark matter halos, the one-halo term\footnote{
The one halo term formula of Eq. (\ref{eq:P1h}) tends to be a constant at large scale (low-$k$), this is an unphysical behaviour \citep{Schaan+2021}. Some attempts have been proposed to solve the problem \citep{Cooray+2002, Baldauf+2013}, but it is still an open issue.} \citep{Moradinezhad-Dizgah2022} is 
\begin{equation}
\begin{aligned}
&P_{\rm CII}^{\rm 1h}(k,z) =\int dM_{\rm vir} \frac{dn}{dM_{\rm vir}} u_{\rm CII}(k|M_{\rm vir})^{2} \left[\frac{L_{\rm CII}(M_{\rm vir},z)}{4\pi D^{2}_{\rm L}} y(z) D_{\rm c}^{2} \right]^{2}\\
&=\left(\frac{c}{4\pi \nu_{\rm CII} H(z)}\right)^{2} \int dM_{\rm vir} \frac{dn}{dM_{\rm vir}} [L_{\rm CII}(M_{\rm vir},z) u_{\rm CII}(k|M_{\rm vir})]^2 ,
\label{eq:P1h}
\end{aligned}
\end{equation}
and two-halo term \citep{Moradinezhad-Dizgah2022} is
\begin{equation}
\begin{aligned}
P_{\rm CII}^{\rm 2h} (k, z) &= \left(\int dM_{\rm vir} \frac{dn}{dM_{\rm vir}} u_{\rm CII}(k|M_{\rm vir})\right)^{2} \left[\frac{L_{\rm CII}(M_{\rm vir},z)}{4\pi D^{2}_{\rm L}} y(z) D_{\rm c}^{2} \right]^{2}\\
& = \left(\int dM_{\rm vir} \frac{dn}{dM_{\rm vir}} L_{\rm CII}(M_{\rm vir},z) u_{\rm CII}(k|M_{\rm vir}) b_{\rm SMT}(M_{\rm vir},z) \right)^{2} \\
& \times \left(\frac{c}{4\pi\nu_{\rm CII}H(z)}\right)^{2} P(k,z).
\label{eq:P2h}
\end{aligned}
\end{equation}
$D_{\rm L}$ is the luminosity distance and $D_{\rm c}$ is the comoving distance to redshift $z$. $y(z)=c(1+z)/[\nu_0H(z)]$ is the derivative of the comoving radial distance with respect to the observed frequency $\nu_0$, which is related to the rest-frame frequency of \CII line as $\nu_0=\nu_{\rm CII}/(1+z)$;
$c$ is the light speed and $H(z)$ is the Hubble parameter.
We adopt the Sheth-Tormen form \citep[][]{Sheth+1999} of the halo mass function $dn/dM_{\rm vir}$, 
the \cite{Eisenstein+1999} form of the linear matter power spectrum $P(k,z)$, and the \cite{Sheth+2001} form of  dark matter halo bias $b_{\rm SMT}(M_{\rm vir},z)$. $L_{\rm CII}(M_{\rm vir}, z) = L_{\rm CII,g}+L_{\rm CII,h}$ is the \CII luminosity of the dark matter halo with mass $M_{\rm vir}$, including the central galaxy and the extended \CII halo.

Finally, the Poissonian shot noise, arising from the discrete nature of dark matter halos, can be computed as \citep{Uzgil+2014}
\begin{equation}
    P_{\rm CII}^{\rm SN}(z) = \left(\frac{c}{4\pi \nu_{\rm CII} H(z)}\right)^{2}        
    \int  d M_{\rm vir}  \frac{dn}{dM_{\rm vir}} L_{\rm CII}^{2}(M_{\rm vir},z)  
    \label{eq:Psn}
\end{equation}

The integration of Eqs. (\ref{eq:P1h}, \ref{eq:P2h}  \&  \ref{eq:Psn}) is performed between $M_{\rm min}$ and $M_{\rm max}$. We set $M_{\rm min}=10^8\,\rm M_\odot$, which is roughly the atomic cooling threshold (the virial mass corresponding to virial temperature  $10^4$ K) at redshift 6. This is the minimum mass that can sustain persistent star formation activity. 
Generally, the number density of dark matter halos with mass $M_{\rm vir}$ can be approximated as $\bar{n}\sim M_{\rm vir}\frac{dn}{dM_{\rm vir}}=\frac{dn}{d\ln M_{\rm vir}}$.
Suppose the survey volume is $V_{\rm  survey}$, then in such volume the mean number of dark matter halos with mass $M_{\rm vir}$ is just 
\begin{equation}
\bar{n}V_{\rm survey}\sim \frac{dn}{d\ln M_{\rm vir}}  V_{\rm survey}.  
\end{equation}

In $V_{\rm survey}$, the probability to find at least one dark matter halo with mass $\sim M_{\rm vir}$ is $G(\ge 1 | \bar{n} V_{\rm survey} )$, where $G$ is the cumulative Poisson probability with mean value $\bar{n}V_{\rm survey}$. We obtain $M_{\rm max}$ by solving the equation $G(\ge 1 | \bar{n} V_{\rm survey} )=0.5$, which means in the volume $V_{\rm survey}$ the contribution from dark matter halos whose existence probability smaller than $ 50\%$ is excluded. This is to avoid such bright and rare objects biasing our results. When $V_{\rm survey}=10^6~$Mpc$^3$, $M_{\rm max}\sim 2.2 \times 10^{12}\,\rm M_{\rm \odot}$ at redshift 6.

Let's interpret the one-halo term, two-halo term, and shot noise of the \CII power spectrum more physically.
Since the mean \CII specific intensity is \citep{Gong+2011}
\begin{equation}
\bar{I}_{\rm CII}(z)= \int dM_{\rm vir} \frac{dn}{dM_{\rm vir}} \left[\frac{L_{\rm CII}(M_{\rm vir},z)}{4\pi D^{2}_{\rm L}} y(z) D_{\rm c}^{2} \right],
\end{equation}
so the one-halo term can be interpreted as the auto-correlations of the \CII light distribution within dark matter halos, weighed by the square of their contribution to the mean \CII specific intensity.
On the other hand, at scales much larger than the dark matter halo size, the two-halo term has the approximation form
\begin{equation}
P_{\rm CII}^{\rm 2h}(k,z)\approx \bar{I}_{\rm CII}^2(z) \bar{b}_{\rm SMT}^2 P(k,z),
\end{equation}
where $\bar{b}_{\rm SMT}$ is the mean halo bias. So the two-halo term actually describes the large-scale power spectrum of the dark matter halos weighted by the square of mean \CII specific intensity. Regarding the shot noise, it has the approximation 
\begin{equation}
P_{\rm CII}^{\rm SN}(z)\approx \bar{I}^2_{\rm CII}(z)\frac{1}{\bar{n}},
\end{equation}
where $\bar{n}$ is the number density of dark matter halos that have a dominant contribution in the shot noise. Clearly, this approximation shows that shot noise is actually the Poisson fluctuations of the dark matter halos, weighted by the square of the mean \CII specific intensity.

\subsection{The signal-to-noise ratio 
 estimation}\label{sec:signal noise ratio}
 
For a LIM survey, the smallest and largest accessible $k$ modes for probing the fluctuations are determined by the survey volume and the resolution (angular resolution $\delta \theta$ and frequency resolution $\delta \nu_0$). The survey volume is determined by the redshift of the target, the survey area $\Omega_{\rm survey}$, and the bandwidth coverage $\Delta \nu_0$.

Along the line of sight, the spatial resolution corresponds to the frequency resolution \citep{Bull2015},
\begin{equation}
    \Delta r_{\rm \parallel, min} \approx \frac{c(1+z)}{H(z)} \frac{\delta \nu_{0}}{\nu_{0}};
\end{equation}
while the largest spatial across corresponds to the bandwidth, 
\begin{equation}
    \Delta r_{\rm \parallel, max} \approx \frac{c(1+z)}{H(z)} \frac{\Delta \nu_{0}}{\nu_{0}}.
\end{equation}

The tangential spatial resolution corresponds to the synthesized beam size $\Omega_{\rm beam}$
\begin{equation}
    \Delta r_{\rm \perp, min} \approx D_{\rm c}(z) \sqrt{\Omega_{\rm beam}},
\end{equation}
and the largest spatial across corresponds to the survey area $\Omega_{\rm survey}$
\begin{equation}
    \Delta r_{\rm \perp, max} \approx D_{\rm c}(z) \sqrt{\Omega_{\rm survey}}.
\end{equation}

Then, along and perpendicular to the line of sight,
the smallest and largest $k$ modes that can be detected by the survey are \citep{Uzgil+2019}, 
\begin{equation}
    k_{\rm \parallel, min} = \frac{2\pi}{\Delta r_{\rm \parallel, max}}\  {\rm and}\  k_{\rm \parallel, max} = \frac{1}{2} \frac{2 \pi}{ \Delta r_{\rm \parallel, min}},
\end{equation}
\begin{equation}
    k_{\rm \perp, min} = \frac{2\pi}{\Delta r_{\rm \perp, max}}\ {\rm and}\ k_{\rm \perp, max} = \frac{1}{2}\frac{2\pi}{\Delta r_{\rm \perp, min}}.
\end{equation}

Finally, the survey detects the isotropic power spectrum in the range $k_{\rm min} < k < k_{\rm max}$, for which $k_{\rm min}\approx \sqrt{k^2_{\parallel,\rm min}+k^2_{\perp,\rm min}}$ and 
$k_{\rm max}\approx \sqrt{k^2_{\parallel,\rm max}+k^2_{\perp,\rm max}}$. 
 
The variance of the power spectrum measured in the above LIM survey is
\begin{equation}
    \sigma_P^{2}(k) = \frac{1}{N_{\rm m}(k)} \left[P_{\rm CII}(k, z)+P_{\rm N}W^{-2}_{\rm res}(k)\right]^{2},
    \label{eq:power spectrum variance}
\end{equation}
where $W_{\rm res}(k)$ is the window function that denotes the rapid decline of the measured power spectrum below the resolution \citep{Lidz+2011, Lidz+2016, Seo+2010, Battye2013, Bernal+2019}. 
In principle $\Delta r_{\parallel,\rm min}$ can be very different from $\Delta r_{\perp, \rm min}$, so the window function for $k_\parallel$ and $k_\perp$  should be different \citep{Lidz+2011, Lidz+2016, Bernal+2019, Bernal+2022}. In that case, at the small scale, the power spectrum is no longer isotropic and the cylinder power spectrum should be used. However, in this paper, we only investigate the isotropic power spectrum, we implicitly assume that the $\Delta r_{\parallel, \rm min}$ are comparable to the $\Delta r_{\perp, \rm min}$. For this reason, we adopt the approximation
\citep{Battye2013}
\begin{equation}
W_{\rm res}(k)\approx \exp\left[-\frac{1}{2}k^2 D^2_{\rm c}(z) \frac{\Omega_{\rm beam}}{4} \right].
\end{equation}  

In principle, there is another window function that describes the power spectrum decline beyond the finite survey volume \citep{Bernal+2019}.
In our calculations we do not consider such a window function, instead, we consider the power spectrum at $k>k_{\rm min}$ is fully measured and at $k < k_{\rm min}$ is fully missed, for which $k_{\rm min}$ is set by the survey volume. 
This is equivalent to adopting a sharp $k$ cutoff step function as the window function. Although this may not be the real case, it will just have quite limited effects on the results, because the contributions to signal-to-noise ratio and parameter constraints from $k$ modes close to $k_{\rm min}$ are small, see the results in Sec.~\ref{sec:detectability}.

In Eq.~(\ref{eq:power spectrum variance}), $P_{\rm N}$ is the instrumental noise power spectrum. $N_{\rm m} (k)$ is the number of $k$ modes sampled by a survey in the $k$-bin, which is estimated \citep{Chung+2020} by \footnote{Note that Eq.~(\ref{eq:k mode number}) may overestimate the number of $k$ modes, particularly at small scales (large $k$) \citep{Gong+2017, Gong+2020}. Because it assumes that all $k$ modes in the survey volume are independent. However, at small scales, $k$ modes are crowded and there must be some degeneracies between them.}
\begin{equation}
\label{eq:k mode number}
    N_{\rm m} (k) \approx 2 \pi k^{3} d \ln{k} \frac{V_{\rm survey}}{(2\pi)^{3}},
\end{equation}
where $d\ln k$ is the relative width of the selected $k$-bin, and $V_{\rm survey}$ is the survey volume, 
\begin{equation}
    V_{\rm survey} \approx D_{\rm c}^{2}(z) \Omega_{\rm survey} \left[\frac{c(1+z)}{H(z)} \frac{\Delta \nu_0}{\nu_{\rm 0}} \right].
\end{equation}
Throughout this paper, we adopt bin width $d \ln k = 0.2$.

Then the signal-to-noise ratio (S/N) of  each $k$-bin
 \begin{align}
\label{eq:signal noise ratio}
    {\rm S/N} &= \frac{P_{\rm CII}(k, z)}{\sigma_P(k)} \nonumber \\
   &= \sqrt{N_{m}(k)} \frac{P_{\rm CII}(k, z)}{P_{\rm CII}(k, z)+P_{\rm N}W_{\rm res}^{-2}(k)},
\end{align}
and the total signal-to-noise is 
\begin{equation}
    (S/N)_{\rm tot} = \left(\sum_i \frac{P_{\rm CII}^{2}(k_i)}{\sigma_P^{2}(k_i)} \right)^{1/2},
\end{equation}
where the sum is performed for all $k$-bins.

If the instrumental noise flux is $\sigma_{\rm N}$, the instrumental noise power spectrum is
\begin{equation}
P_{\rm N}=\left(\frac{\sigma_{\rm N}}{\Omega_{\rm beam}}\right)^2V_{\rm vox},
\label{eq:P_N}
\end{equation}
where 
\begin{equation}
    V_{\rm vox} \approx D_{\rm c}^{2}(z) \Omega_{\rm beam} \Big[\frac{c(1+z)}{H(z)} \frac{\delta \nu_0}{\nu_0}\Big].
\end{equation}
is the comoving volume of the real space voxel.

Suppose the survey is carried out by an interferometer array  with $N_{\rm an}$ antennas, each antenna has diameter $d$ and system temperature $T_{\rm sys}$, then the instrumental noise flux
\begin{equation}
\sigma_{\rm N}=\frac{2 k_{\rm B} T_{\rm sys}} {d^2\sqrt{N_{\rm an}(N_{\rm an}-1)/2 \delta \nu_0t_{\rm int}}}, 
\end{equation}
where $t_{\rm int}$  is the integration time on source. 
Suppose each antenna has a primary beam 
\begin{equation}
\Omega_{\rm primary}\sim\left(\frac{\lambda_{\rm obs}}{d}\right)^2,
\end{equation}
and $N_{\rm chn}$ frequency channels. Then for each instantaneous pointing the array can survey a fraction $f_{\rm survey}=\frac{\Omega_{\rm primary}}{\Omega_{\rm survey}} \frac{ N_{\rm chn} \delta \nu_0} {\Delta \nu_0} $ ~($N_{\rm chn} \delta \nu_0 \le \Delta \nu_0$) of the target volume.
Therefore 
\begin{align}
t_{\rm int}&=t_{\rm obs} f_{\rm survey} \nonumber \\
&=t_{\rm obs} \frac{\Omega_{\rm primary}}{\Omega_{\rm survey}} \frac{ N_{\rm chn} \delta \nu_0} {\Delta \nu_0}~~~~~(N_{\rm chn}\delta \nu_0 \le \Delta \nu_0), 
\end{align}
where $t_{\rm obs}$ is the total observation time.

If the interferometer has maximum baseline length $b_{\rm max}$, then
the Full Width at Half Maximum (FWHM) of the beam is $\theta_{\rm FWHM}=1.2\lambda_{\rm obs}/b_{\rm max}$, for Gaussian profile $\theta_{\rm beam}=\theta_{\rm FWHM}/\sqrt{2\ln2}$, hence $\Omega_{\rm beam}=(\theta_{\rm beam})^2=(\theta_{\rm FWHM}/\sqrt{2\ln 2})^2$.

\section{Results}\label{sec:results}

\subsection{\CII power spectrum }

\begin{figure*}
	\includegraphics[width=2\columnwidth]{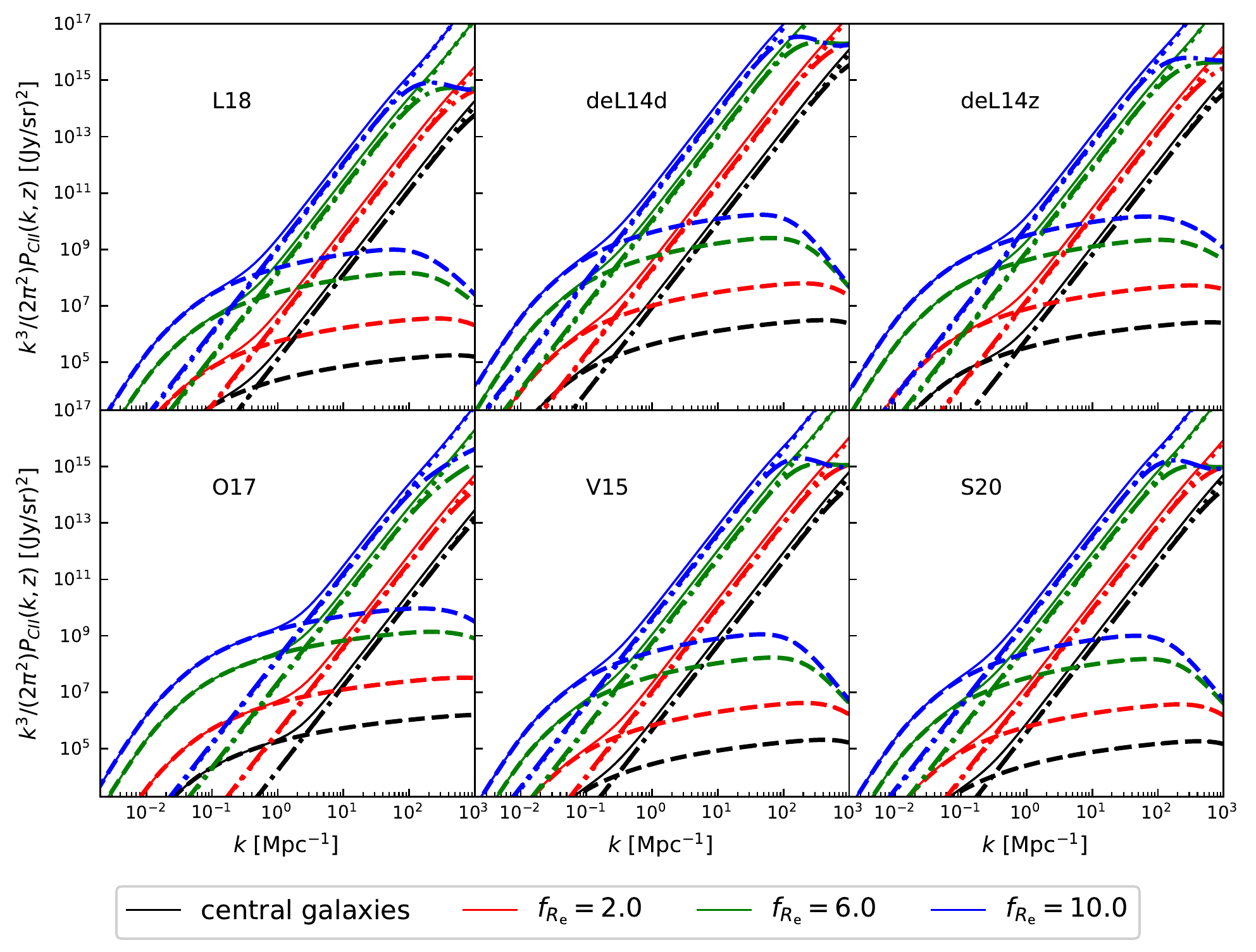}
    \caption{Predicted \CII power spectrum for six \CII-SFR relations. The dash-dotted line, dashed line, dotted line, and thin solid line denote the one-halo, two-halo, shot noise terms, and the sum of them respectively, in both panels. The dotted line (shot noise) overlaps with the dash-dotted line (one-halo term) at a large scale. The black line represents the \CII power spectrum for the central galaxy, i.e. without \CII halo considered. Red, green, and blue lines corresponding to $f_{R_{\rm e}}=[2.0, 6.0, 10.0]$, with fixed $f_{\Sigma} \sim 0.4$.}
    \label{fig:signal different f_Re}
\end{figure*}

\begin{figure*}
	\includegraphics[width=2\columnwidth]{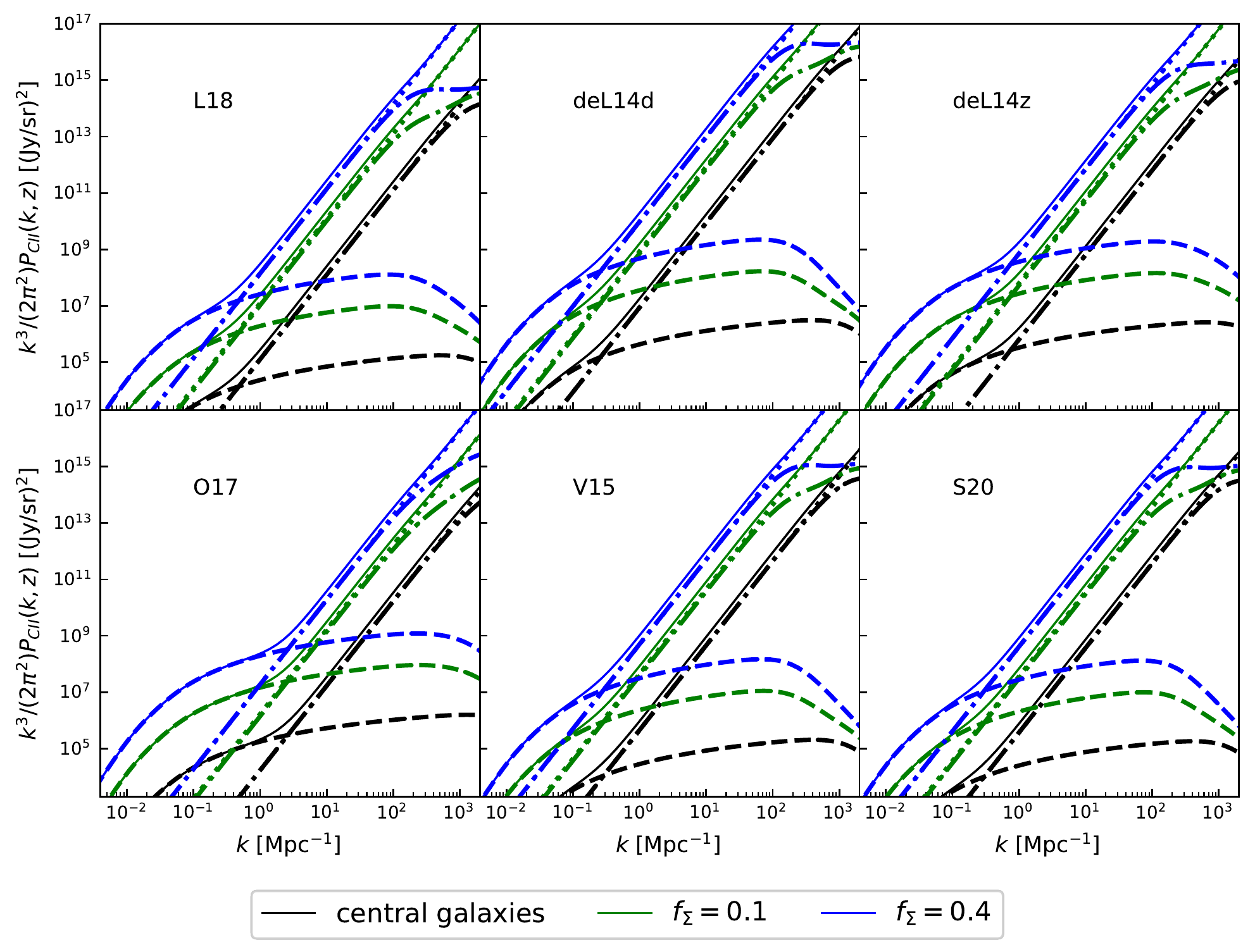}
    \caption{As in Fig. \ref{fig:signal different f_Re} for different values of $f_{\Sigma}$ and for the six different \CII-SFR relations considered (see Tab. \ref{tab:CII SFR relation}); we fix $f_{R_{\rm e}} = 6.0$ here.}
    \label{fig:signal different f_sigma}
\end{figure*}

In Fig.~\ref{fig:signal different f_Re}, we plot the intensity mapping power spectrum at $z \sim 6$ for the six $L_{\rm SFR, CII}-\rm SFR$ relations with effective radius ratio values $f_{R_{\rm e}}=[2.0, 6.0, 10.0]$, representing conservative, moderate, and extreme cases, respectively. We fix $f_{\Sigma}=0.4$, i.e. the central surface brightness of the halo represents $\sim 30\%$ of the total, which is a reasonable guess from the fitting results of the observational data \citep{Akins+2022}. It clearly shows that \CII halos largely boost the intensity of the power spectrum. Compared with the power spectrum of central galaxies, 
the signal is boosted by
$\sim 20, 10^{3}, 10^{4}$ 
times when $f_{R_{\rm e}} = 2.0, 6.0, 10.0$ independently on the specific form of $L_{\rm CII, SFR}-\rm SFR$ relations. Moreover, \CII halos imprint a specific signature in the one-halo term at small scales. 
Since the central galaxy is more compact, the typical turnover scale of the one-halo term is $\sim10^3$ Mpc$^{-1}$. However, the \CII halo is more extended and the typical turnover scale is shifted to $\sim100$ Mpc$^{-1}$. 
However, this feature is generally buried in the shot noise, which is the generally dominant component at the smallest scales.
Fig.~\ref{fig:signal different f_sigma} illustrates
instead the power spectrum dependence on $f_{\Sigma}$ at a fixed value of $f_{R_{\rm e}}=6.0$. We show the results for $f_{\Sigma}=0.1, 0.4$, where the signal is strengthened by $\sim 10^{2}, 10^{3}$ times, respectively. 

\begin{figure}
	\includegraphics[width=\columnwidth]{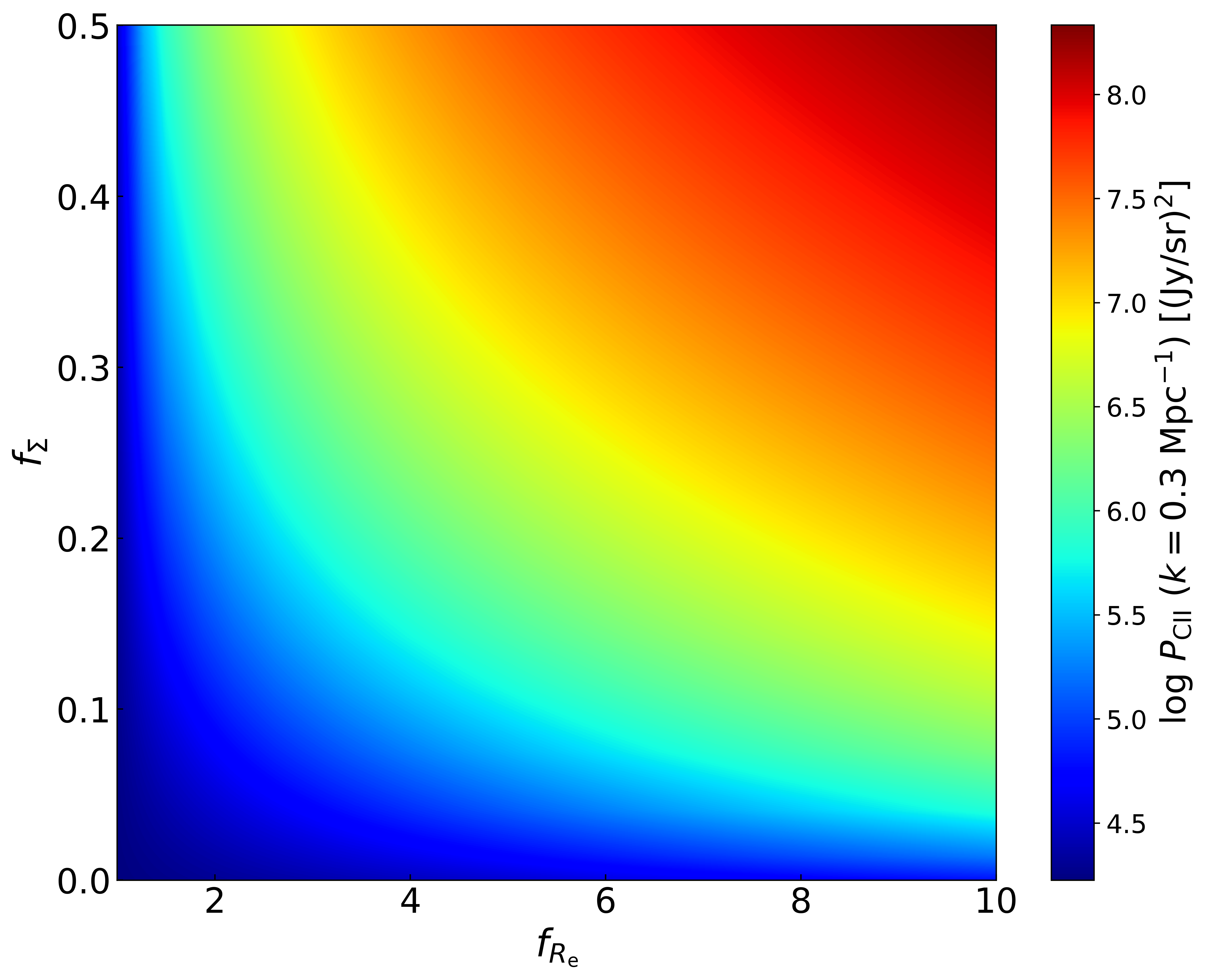}
    \caption{The total \CII halo power spectrum at $k=0.3\ {\rm Mpc}^{-1}$ as a function of $f_{\Sigma}$ and $f_{R_{\rm e}}$, to avoid repetition we only show the result for L18 \CII-SFR relation.}
    \label{fig:parameter_space}
\end{figure}

We also explore the results in the 
$f_{R_{\rm e}}-f_\Sigma$
parameter space, as shown in Fig.~\ref{fig:parameter_space}. We plot the values of the total power spectrum, $P_{\rm CII}$, at $k = 0.3 \,{\rm Mpc}^{-1}$ by varying $f_{\Sigma}$ and $f_{R_{\rm e}}$ simultaneously. To avoid repetition, we only show the result for the L18 relation. As we can see, $P_{\rm CII}$ is sensitive to both changes in $f_{R_{\rm e}}$ and in $f_{\Sigma}$.

\subsection{Detectability of the \CII halo signal}\label{sec:detectability}

We now turn the discussion to the detectability of the predicted \CII power spectrum signal in a LIM survey.

\begin{subsubsection}{Detectability in ALMA ASPECS Survey}

The ALMA Large Program ASPECS \citep[][]{Aravena+2020} surveying the Hubble Ultra Deep Field (HUDF) provides the first full frequency scan in Band 6, corresponding to the frequency window for \CII emission from $6<z<8$ galaxies, we first consider the detectability of \CII power spectrum signal by such an experiment.  

The ASPECS Band 6 data covers a total area of $4.2\ \rm arcmin^{2}$ in the HUDF, with $2.9\ \rm arcmin^{2}$ area within the 50\% primary beam response \citep[][]{Decarli+2020}. The observed frequency range is $212-272\ \rm GHz$, corresponding to the redshift range $z = 5.99-7.97$ for the \CII line. If we rebin the frequency channels by a factor of 8, as suggested by \citet[][]{Uzgil+2021}, the spectral resolution is $\delta \nu_0 = 62.5\, \rm MHz$. The synthesized beam size (the FWHMs of the pixel ellipse along its major and minor axes, see \citealt{Uzgil+2019}) in the image cube is  $ \Delta \theta_{\rm b, maj} \times \Delta \theta_{\rm b, min} =  1.6'' \times 1.1'' $ \citep[][]{Uzgil+2021}. Then
\begin{equation}
    \Omega_{\rm beam} = \frac{ \theta_{\rm b, maj} \times \theta_{\rm b, min}}{2 \ln 2}. 
\end{equation}

From these survey parameters, we derive 
$k_{\parallel,\rm min}=0.009\, \rm Mpc^{-1}$,
$k_{\parallel, \rm max}=4.551\, \rm Mpc^{-1}$,
$k_{\perp,\rm min}= 1.503\, \rm Mpc^{-1}$, and
$k_{\perp,\rm max}= 68.147\, \rm Mpc^{-1}$, and we finally have $k_{\rm min}= 1.503\, \rm Mpc^{-1}$, $k_{\rm max}= 68.299\, \rm Mpc^{-1}$. We summarize the ASPECS survey parameters derived for the \CII power spectrum analysis in Tab.~\ref{tab:ASPECS_parameters}.

Next, we compute the S/N of the \CII power spectrum for an ASPECS-like survey. The survey has noise flux $\sigma_{\rm N}  = 0.30\,\rm mJy\,\rm beam^{-1}$ \citep[][]{Uzgil+2021}, which yields a surface brightness intensity sensitivity $\sigma_{\rm N}/\Omega_{\rm beam}=1.01 \times 10^{7}\, \rm Jy\ sr^{-1} $, and a noise power spectrum $P_{\rm N} = 1.48 \times 10^{11}\ \rm (Jr/sr)^{2}\ \,{{\rm Mpc}}^{3}$ using Eq. (\ref{eq:P_N}).

We show the results of S/N for various $L_{\rm CII,SFR}-{\rm SFR}$ relations with $f_\Sigma=0.4$, $f_{R_{\rm e}}=2.0$ (left panel, represents a relatively compact \CII halo) and $f_\Sigma=0.1$, $f_{R_{\rm e}}=6.0$ (right panel, represents a relatively diffuse \CII halo) respectively in Fig.~\ref{fig:S/N_ASPECS}. In the left panel, only the signal of the deL14d \CII - SFR model is detectable (with ${\rm  S/N}\gtrsim3$) in the range $10\,\rm Mpc^{-1} \lesssim k \lesssim 70\,\rm Mpc^{-1}$. In the right panel, the model deL14d is detectable in $4\,\rm Mpc^{-1} \lesssim k \lesssim 70\,\rm Mpc^{-1}$,  while the model deL14$z$ is detectable in $10\,\rm Mpc^{-1} \lesssim k \lesssim 70\,\rm Mpc^{-1}$. This is because although the surface brightness of the \CII halo is lower in the right panel, the total luminosity is larger as it scales with $f_{R_{\rm e}}^{2}$. We additionally demonstrate the impact of the shot noise on S/N by dotted lines. When shot noise is excluded, the resulting S/N is reduced by a factor of $\sim 2$. This is because the ASPECS has a small area. The detected power spectrum is basically the one-halo clustering term plus the shot noise, the two-halo clustering term is almost negligible. While the shot noise is comparable to the one-halo clustering term, excluding it will reduce the shot noise by a factor $\sim2$.

\begin{table}
\begin{center}
\caption{The ASPECS survey parameters for S/N analysis.}
\label{tab:ASPECS_parameters}
\begin{tabular}{  l r  }
\hline
\hline
 [$\nu_0-\Delta \nu_0,\nu_0$] & [212—272]\ GHz  \\ 
 $z$ & [5.99, 7.97] \\
$\delta \nu_0$ & 62.5\ MHz  \\
 $\Omega_{\rm survey}$ & $2.9\,\rm arcmin^{2}$ \\
 $\Omega_{\rm beam}$ & $1.6'' \times 1.1''$ \\
 $\sigma_{\rm N}$ & 0.30 mJy beam$^{-1}$ \\
 \hline

$V_{\rm survey}$& 13260 Mpc$^3$ \\

$\Delta r_{\parallel,\rm min}$ & 0.690 Mpc \\

$\Delta r_{\parallel,\rm max}$ & 662.701 Mpc \\

$\Delta r_{\perp,\rm min}$ & 0.046 Mpc \\

$\Delta r_{\perp,\rm max}$ & 4.180 Mpc \\

$k_{\parallel,\rm min}$ & 0.009 Mpc$^{-1}$ \\

$k_{\parallel,\rm max}$ & 4.551 Mpc$^{-1}$ \\

$k_{\perp,\rm min}$ & 1.503 Mpc$^{-1}$ \\

$k_{\perp,\rm max}$ & 68.299 Mpc$^{-1}$ \\

$k_{\rm min}$ & 1.503 Mpc$^{-1}$ \\

$k_{\rm max}$ & 68.299  Mpc$^{-1}$ \\

$P_{\rm  N}$ & $1.48\times10^{11}$ (Jy/sr)$^2$  Mpc$^3$ \\

\hline
\end{tabular}
\end{center}
\end{table}

\begin{figure*}
	\includegraphics[width=2\columnwidth]{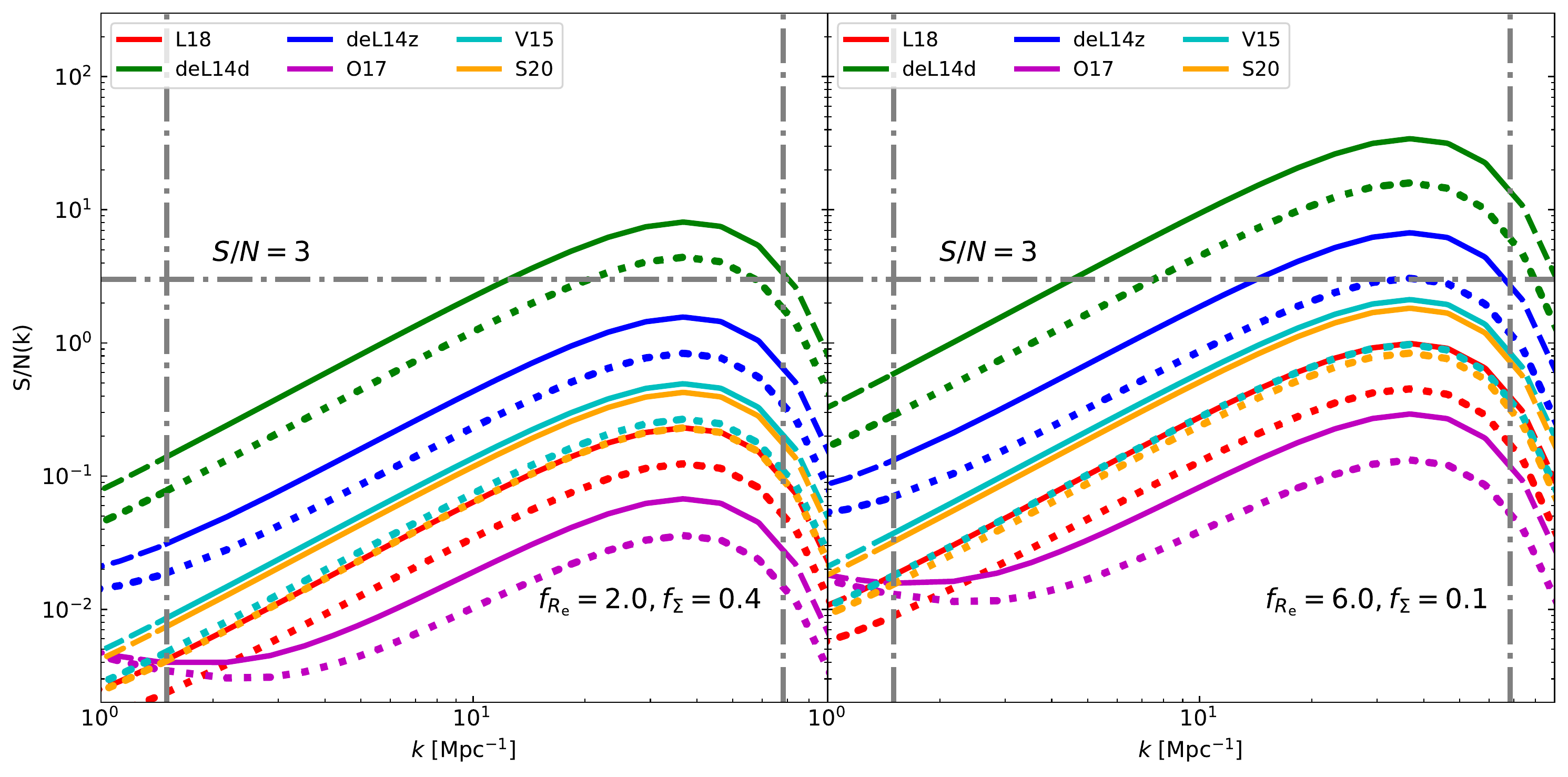}
    \caption{S/N for six different $L_{\rm CII, SFR}-\rm SFR$ relations for the ALMA ASPECS Survey. The solid lines represent the S/N for the total power spectrum (including contributions from one-halo and two-halo terms as well as shot noise), while the dotted lines illustrate the S/N when the shot noise is excluded. The $k_{\rm min}$ and $k_{\rm max}$ are shown as grey dashed-dotted vertical lines. \textit{Left panel:} case with $f_{R_{\rm e}} = 2.0, f_{\Sigma}=0.4$. \textit{Right panel:} same for $f_{R_{\rm e}}= 6.0, f_{\Sigma}=0.1$.}
    \label{fig:S/N_ASPECS}
\end{figure*}

\end{subsubsection}

\begin{subsubsection}{Optimal survey strategy}
To optimally probe the extended \CII halo signal, an ideal survey should be able to detect the signal with a resolution up to scale comparable to the extended \CII halo size. On the other hand, to enhance the statistical significance of the signal, the survey should cover a sky area much larger than the ASPECS. Here we propose, by using ALMA 12-m antennas in an extended configuration with $\sim 1000$ m baseline, $N_{\rm an} = 500$, and $N_{\rm chn}=1000$,  to survey a total area of $\Omega_{\rm survey} \sim 2\,\rm deg^{2}$ in frequency band [212-272] GHz with total observing time $t_{\rm obs} \sim 1000\,\rm hr$. The frequency resolution $\delta \nu_0=\Delta \nu_0/N_{\rm chn}=60$ MHz.

The system temperature\footnote{$T_{\rm sys}$ comes from the \href{https://almascience.eso.org/proposing/sensitivity-calculator}{ALMA Sensitivity Calculator (ASC)} by setting the observing frequency $\nu_0 = 272\,\rm GHz$ and bandwidth $\Delta \nu_0 = 60\, \rm GHz$.}
of ALMA at observing frequency $\nu_{0} = 272\,\rm GHz$ is $T_{\rm sys} \sim 115\,\rm K$. Following the procedures in Section~\ref{sec:signal noise ratio}, we calculate the S/N and show them in Fig.~\ref{fig:S/N_optimal}. 
A summary of the parameters for this designed survey is given in Tab. \ref{tab:deisgned_parameters}.

For this optimized survey, when $f_\Sigma=0.4$ and $f_{R_{\rm e}}=2.0$ (see the left panel of Fig. \ref{fig:S/N_optimal}), the \CII power spectra of the six $L_{\rm CII, SFR}-\rm SFR$ relations all detectable with total signal-to-noise ratio $\gtrsim10$. For the more extended \CII halo model with $f_\Sigma=0.1$ and $f_{R_{\rm e}}=6.0$ (see the right panel of Fig. \ref{fig:S/N_optimal}), the total signal-to-noise ratios of the power spectral are even higher. Similar to Fig.~\ref{fig:S/N_ASPECS}, we show the S/N if the shot noise is not involved by dotted lines.  
Different from the Fig. \ref{fig:S/N_ASPECS}, when shot noise is excluded, the S/N does not change at $k\lesssim 0.1$ Mpc$^{-1}$. However, at $k\gtrsim 0.1$ Mpc$^{-1}$ where the shot noise dominates over the clustering term, the S/N is reduced by a factor $\sim 2$, similar to Fig. \ref{fig:S/N_ASPECS}. This is natural since at $k\gtrsim 0.1$ Mpc$^{-1}$, the shot noise term is comparable to the one-halo clustering term. 

\begin{figure*}
	\includegraphics[width=2\columnwidth]{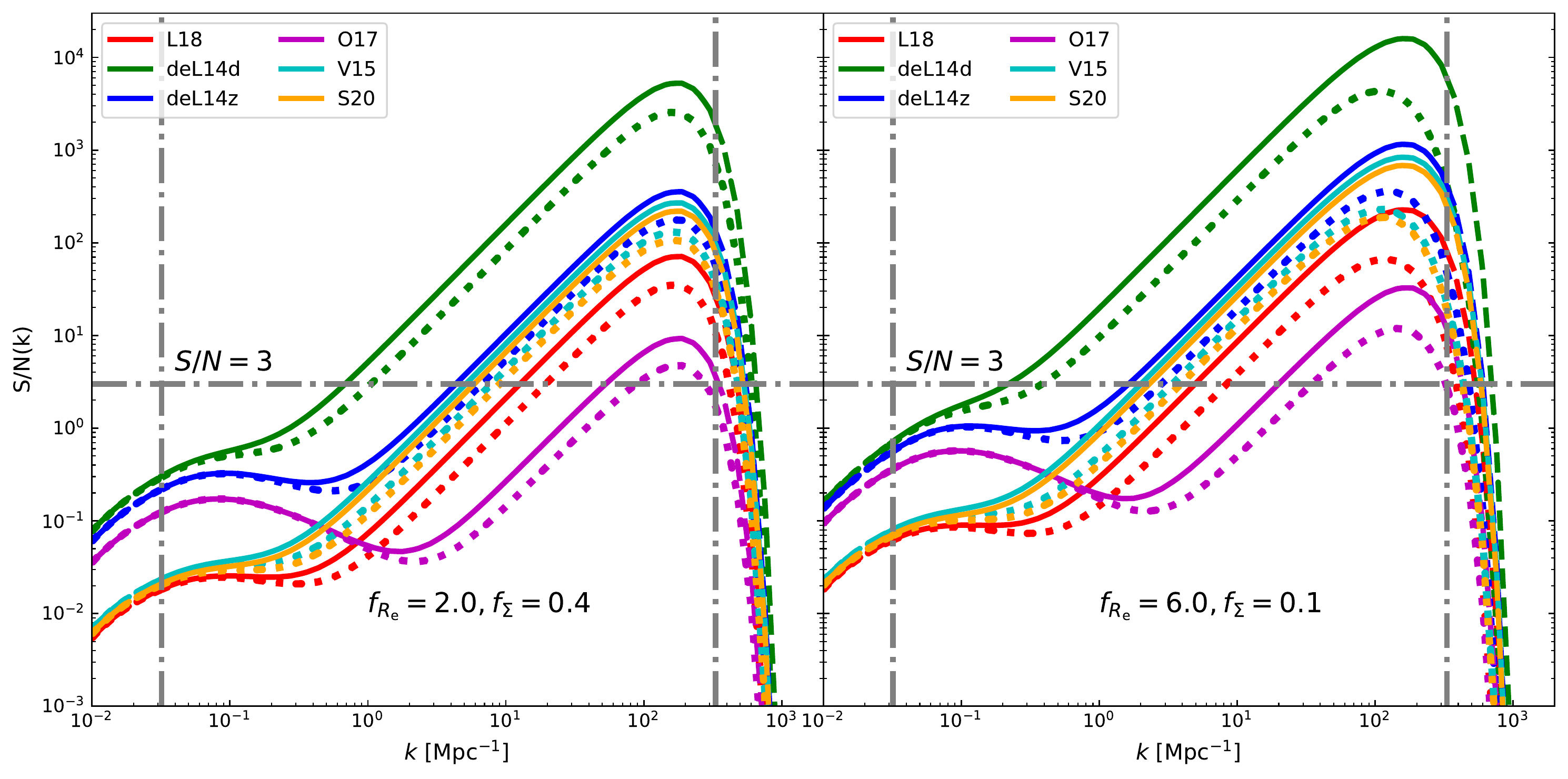}
    \caption{As in Fig. \ref{fig:S/N_ASPECS} but for the optimal survey strategy by using ALMA. With $1000\,\rm m$ baseline, the largest $k$ mode can be probed is $k_{\rm max}=331.43 \,{\rm Mpc}^{-1}$. The signal for the six $L_{\rm CII, SFR}-\rm SFR$ relations can be detected by this configuration in both two cases.}
    \label{fig:S/N_optimal}
\end{figure*}

\begin{table}
\begin{center}
\caption{Parameters of a  designed optimal survey
}
\label{tab:deisgned_parameters}
\begin{tabular}{  l r }
\hline
\hline
$d$ & 12 m \\
 
$T_{\rm sys}$ & 115  K \\

$N_{\rm an}$  & 500  \\

$N_{\rm chn}$  & 1000 \\

$b_{\rm max}$ & 1000 m  \\
 $\theta_{\rm beam}$ & 0.232$''$  \\
 $[\nu_0-\Delta  \nu_0,\nu_0]$ & [212,272]  GHz  \\ 
 Redshift range & [6.0-8.0]  \\
  $\delta \nu_0$ & 60  MHz  \\

$t_{\rm obs}$ & 1000  hr \\

$\Omega_{\rm survey}$  & 2  $\rm deg^{2}$ \\ 
 $V_{\rm survey}$ &$2.88 \times 10^{7}$  $\rm Mpc^{3}$ \\
 $V_{\rm vox}$ & $5.96 \times 10^{-5}$  $\rm Mpc^{3}$ \\
 $P_{\rm N}$ & $ 4.87 \times 10^{11}$  $\rm (Jy\ sr^{-1})^{2}\ Mpc^{3}$ \\

$k_{\rm min}$  & 0.032  $\rm Mpc^{-1}$ \\
 $k_{\rm max}$ & 331.43  $\rm Mpc^{-1}$ \\

 \hline
\end{tabular}
\end{center}
\end{table}

\end{subsubsection}

\begin{subsection}{Constraining the {\rm \CII} halo parameters}

We have shown that when the contribution from extended \CII halos is considered, the \CII power spectrum is boosted. However, this is mainly because the total \CII luminosity of the system (central galaxy + \CII halo) is larger. So there is a degeneracy between the \CII halo contribution and the \CII - SFR relation.  \CII halos also change the shape of the \CII power spectrum (one-halo term) at scales comparable to their size. However, the scales are so small, that generally, the shot noise is much larger than the one-halo term. Here we investigate whether our designed optimal survey is able to put some conclusive constraints on the \CII  halo properties. 

We first generate the mock observed power spectrum $P_{\rm CII}^{\rm obs}$ for our designed optimal survey from Eq. (\ref{eq:P_CII}), by adopting $\log  A= 6.5$, $\gamma=1.2$, $f_\Sigma=0.1$ and $f_{R_{\rm e}}=6.0$ as the input \CII halo parameters. 
Since in this paper we focus on constraining the \CII halo parameters, we fix the cosmological parameters to reduce the freedoms. We adopt the cosmological parameters \citealt{Planck2016}.
The mock uncertainties are from Eq. (\ref{eq:power spectrum variance}).  We then obtain the forecast on constraining the \CII halo parameters by minimizing
\begin{equation}
    \chi^{2} (\Theta) = \sum_{ i=1}^{N_{\rm bin}} \frac{\left[P_{\rm CII}(k_{i}, \Theta)-P_{\rm CII}^{\rm obs}(k_{i})\right]^{2}}{\sigma_P^{2}(k_{i})}
    \label{eq:Chi_square}
\end{equation}
using the MCMC procedure \citep{Goodman+2010}, where $\Theta=\{\log A, \gamma, f_\Sigma, f_{R_{\rm e}}\}$ represents the $L_{\rm CII,SFR}-{\rm SFR}$ relation and extended \CII halo parameters; $N_{\rm bin}$ is the number of $k$ bins between $k_{\rm min}$ and $k_{\rm max}$, with width $d\ln k=0.2$. When performing the MCMC procedure, we set flat priors $\log A \in [5.0, 9.0], \gamma \in [0.1, 2.0], f_\Sigma \in [0.0, 0.5], f_{R_{\rm e}} \in [1.0, 10.0]$. 

In Fig. \ref{fig:mock_data}, we show the mock observed power spectrum with uncertainties and the best-fitting cure obtained from our MCMC fitting. The forecast on the constraints of \CII halo parameters for our designed optimal survey is shown in Fig. \ref{fig:mcmc_constrain_optimal}. The  marginalized parameters for \CII halo are: $f_\Sigma= 0.102 ^{+0.015}_{-0.018}$, $f_{R_{\rm e}}=5.894 ^{+0.912}_{-0.693}$. Indeed, it is feasible to distinguish the extended \CII halo contribution to the power spectrum in this designed optimal survey. However, we also check that the ASPECS survey is still hard to give conclusive constraints, because of the smaller S/N and lower resolution. 

We also perform a MCMC fitting for which we set the shot noise as an independent free parameter, instead of a function of $\Theta$. We obtain $f_\Sigma=0.216^{+0.152}_{-0.109}$ and $f_{R_{\rm e}}=5.917_{-0.686}^{+0.683}$. The constraints become looser. 
This is not surprising, because treating the shot noise as an independent parameter increases the degree of freedom. Moreover, in our case, the one-halo term contributes most to the constraints on \CII halo parameters because it contains the information of \CII luminosity profile. However, the shot noise, if expressed as a function of $\Theta$, also provides some constraints on $\Theta$. Shot noise is also a kind of useful signal, although using this term solely will suffer from heavy parameter degeneracy.

\begin{figure}
	\includegraphics[width=\columnwidth]{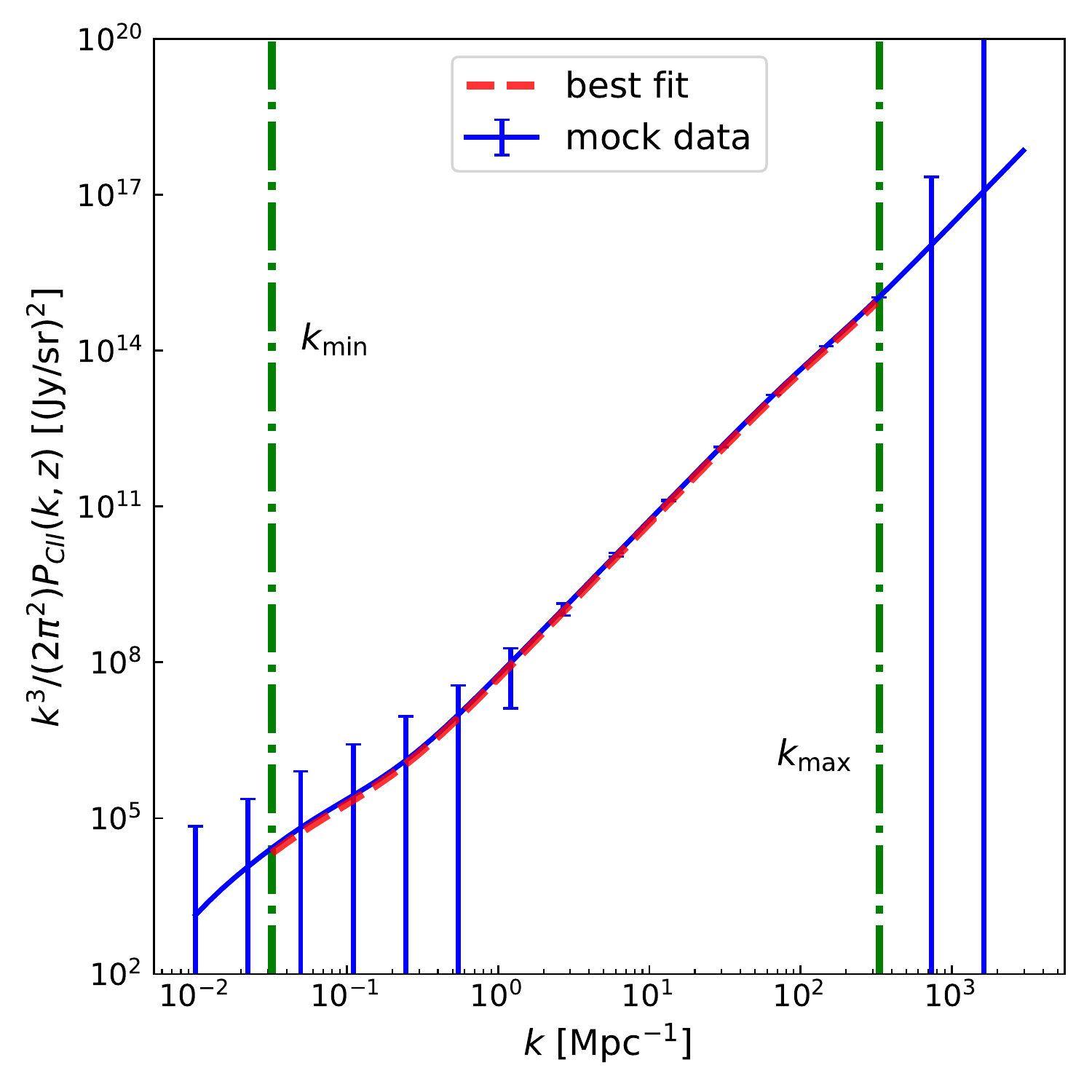}
    \caption{The mock observed power spectrum $P_{\rm CII}^{\rm obs}(k)$ (solid line with errorbars) and the best fitting power spectrum in $k_{\rm min}<k<k_{\rm max}$ (dashed line). We mark $k_{\rm min}$ and $k_{\rm max}$ by two vertical dashed-dotted lines. }
\label{fig:mock_data}
\end{figure}

\begin{figure*}
	\includegraphics[width=2\columnwidth]{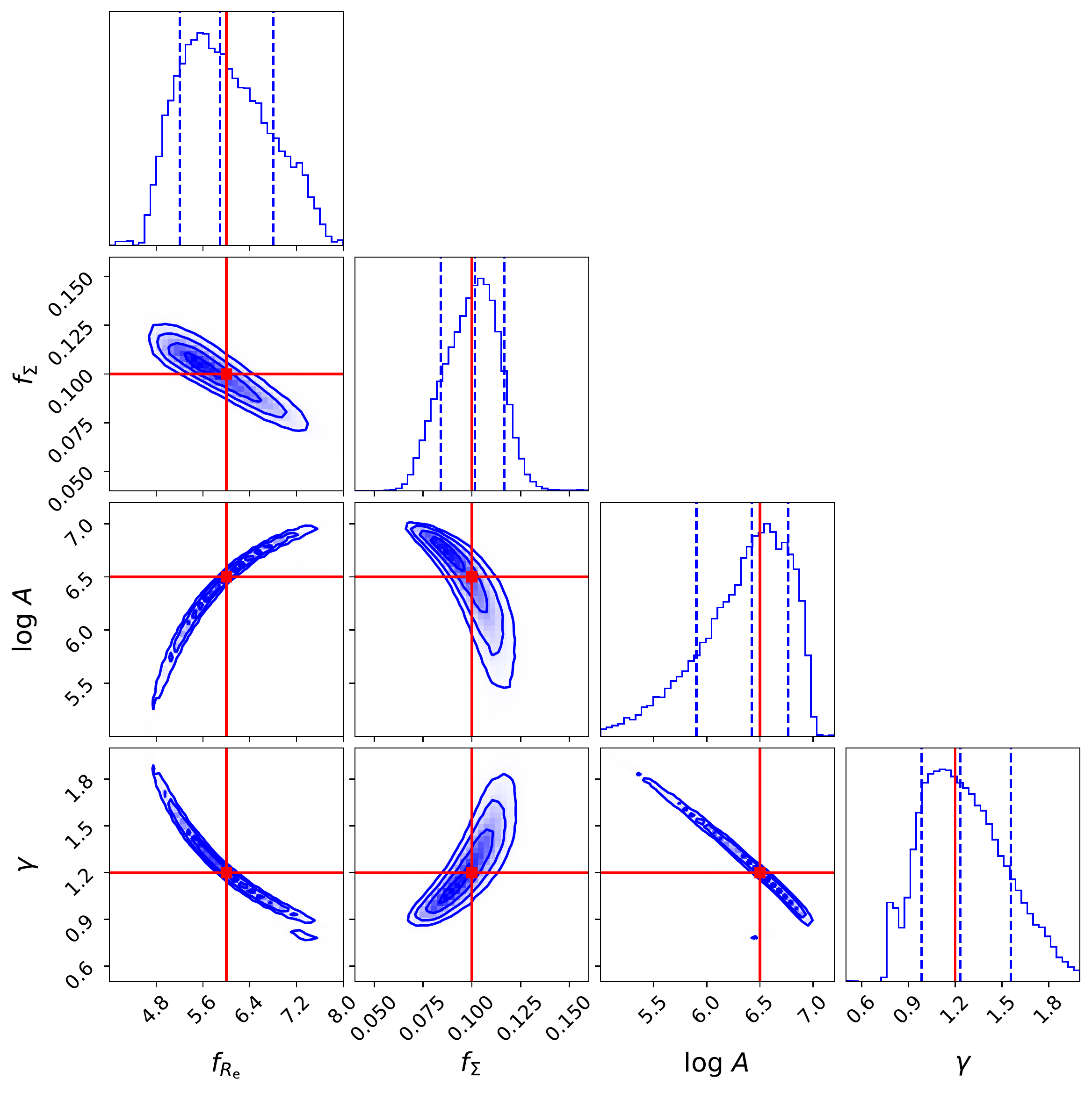}
    \caption{The forecast of confidence levels  of the $L_{\rm CII,SFR}-{\rm SFR}$ relation and \CII halo parameters obtained by the optimized survey. We mark the input parameters by the crosses of vertical and horizontal lines.}
\label{fig:mcmc_constrain_optimal}
\end{figure*}

\end{subsection}

\section{Conclusions}\label{sec:conclusion}

In this paper, we predicted the foreground- and contamination-cleaned \CII power spectrum signal at $z \approx 6$, when both central galaxies and extended \CII halos are considered, and investigate the detectability of such signal using ALMA ASPECS survey and a designed optimized survey. We modelled the \CII  luminosity profiles of the central galaxy and extended \CII halo by a $\rm S\acute{e}rsic$+exponential profile, and derive the \CII- dark matter halo mass relation by matching the dark matter halo mass function with dust-corrected UV luminosity function of high-redshift galaxies. The main results are:

\begin{itemize}

\item The extended \CII halos around high redshift galaxies can significantly enhance the LIM signal compared with the signal produced by central galaxies alone, both in terms of the clustering signal and shot noise. 

\item Our \CII halo model has two free parameters: the effective \CII halo/galaxy radius ratio, $f_{R_{\rm e}}$, and the central surface brightness ratio, $f_{\Sigma}$. The luminosity of extended \CII halos is $\sim 2.24 f_{\Sigma} f_{R_{\rm e}}^{2}$ times the central galaxies. When the \CII halo contribution is included, compared with the power spectrum from central galaxies only, the signal is boosted by a factor from $\sim 20$ to $\sim 10^{3}$ when $f_{R_{\rm e}}$ varies between $2.0$ to $6.0$,  if $f_{\Sigma}=0.4$.  Given $f_{R_{\rm e}}=6.0$, the power spectrum is enhanced by $\sim 100$ to $\sim 10^{3}$ times when $f_{\Sigma}$ changes from 0.1 to 0.4.

\item For a LIM experiment configured as the ALMA ASPECS Large Program (with resolution $\theta_{\rm beam}=1.13''$ and survey area $\Omega_{\rm survey}=2.9\,\rm arcmin^{2}$), the \CII power spectrum signal is detectable (S/N$\gtrsim3$) for the deL14d  $L_{\rm CII, SFR}-{\rm SFR}$ relation with $f_\Sigma=0.4$ and $f_{R_{\rm e}}=2.0$, and for the deL14d, deL14$z$ relations with $f_\Sigma=0.1$ and  $f_{R_{\rm e}}=6.0$.
So for a LIM experiment, the signal of more extended, low surface brightness \CII halos is more easily detected.

\item To optimally detect the signal, we proposed an optimized survey using more ALMA antennas and longer baselines. The survey has a higher resolution ($\theta_{\rm beam} \sim 0.232''$) and larger survey area ($\Omega_{\rm survey} \sim 2\,\rm deg^{2}$). The resulting signal-to-noise ratio is large enough so that the signal from six $L_{\rm CII, SFR}-{\rm SFR}$ relations is detectable. We also predicted the confidence levels of the constraints on the  $L_{\rm CII, SFR}-{\rm SFR}$ relation and \CII halo parameters by this optimized survey. These two halo parameters are degenerate, but we still expect to obtain meaningful constraints, with uncertainties of $\sim 15\%$ on $f_\Sigma$ and $\sim 10\%$ on $f_{R_{\rm e}}$. 
\end{itemize}
 
In this work, we ignored the continuum foreground and the interloping lines, assuming they could be ideally removed and a clean \CII signal plus pure instrumental white noise is obtained. In practice, however, this is quite a challenge. The IR continuum foreground from dust emission of extragalactic star-forming galaxies can be $\gtrsim10^3$ times higher than the \CII LIM signal (e.g. \citealt{Yue+2015, Bethermin+2022, Cuyck+2023}), while the interloping lines (for example the CO and \CI lines from low redshift galaxies) can also be much higher than the \CII line signal (e.g. \citealt{Yue+2019, Bethermin+2022, Cuyck+2023}). 
The Galactic dust emission and CMB also play a crucial role \citep{Yue+2019}. In our model, the \CII LIM signal is enhanced because of the contribution from \CII halos. However, it is still far from enough to overcome the foreground/intervening lines. 
Many methods have been proposed to remove the foreground and interloping lines. We list some of them in Appendix~\ref{sec:Contamination removal}. We refer interested readers to the references therein.

To show the influence of foreground and interloping lines, one can make mock observations by simulations, and then apply the removal algorithms. However, it is beyond the scope of this work.  Generally, if there are some residual foreground/interloping lines, particularly at a small scale, our constraints on the \CII halo parameters will be biased. 

Our predictions depend on the effective radius ratio, the central surface brightness ratio, and the choice of the $L_{\rm CII, SFR}-\rm SFR$ relation. Therefore, detection of the \CII halo signal in the LIM power spectrum can be used to verify whether \CII halos are ubiquitous in high-redshift galaxies, and provide information about \CII halo size, and their possible relation with outflows carrying the emitting material out of the main galaxy body. Finally, it can be potentially used to constrain the $L_{\rm CII, SFR}-\rm SFR$ relation at high-$z$ as the amplitude of the power spectrum is sensitive to such quantity.

\section*{Acknowledgements}
We thank the anonymous referee for the useful comments that helped us to improve this paper.
MZ acknowledges the financial support from the China Scholarship Council (CSC, No.202104910322). AF acknowledges support from the ERC Advanced Grant INTERSTELLAR H2020/740120. Partial support from the Carl Friedrich von Siemens-Forschungspreis 
der Alexander von Humboldt-Stiftung Research Award is kindly acknowledged. 
BY acknowledges the support by the National SKA Program of China No. 2020SKA0110402. We gratefully acknowledge the computational resources of the Center for High Performance Computing (CHPC) at SNS. We make use of Scipy \citep{Scipy}, Numpy \citep{numpy}, Matplotlib \citep{Hunter2007}, emcee \citep{Foreman-Mackey+2013}, and corner \citep{corner} package for Python to do the calculations and produce the plots.

\section*{Data Availability}

The data produced in this study are available from the corresponding author upon reasonable request.

%%%%%%%%%%%%%%%%%%%% REFERENCES %%%%%%%%%%%%%%%%%%

% The best way to enter references is to use BibTeX:

\bibliographystyle{mnras}
\bibliography{references} % if your bibtex file is called example.bib

% Alternatively you could enter them by hand, like this:
% This method is tedious and prone to error if you have lots of references
%\begin{thebibliography}{99}
%\bibitem[\protect\citeauthoryear{Author}{2012}]{Author2012}
%Author A.~N., 2013, Journal of Improbable Astronomy, 1, 1
%\bibitem[\protect\citeauthoryear{Others}{2013}]{Others2013}
%Others S., 2012, Journal of Interesting Stuff, 17, 198
%\end{thebibliography}

%%%%%%%%%%%%%%%%%%%%%%%%%%%%%%%%%%%%%%%%%%%%%%%%%%

%%%%%%%%%%%%%%%%% APPENDICES %%%%%%%%%%%%%%%%%%%%%

\appendix

\section{Profile deprojection}\label{sec:deprojection}

The deprojection of the 2D surface density profile $\Sigma(R)$ is the inversion of the Abel integral \citep{Binney+1982, Binney+1987}  
\begin{equation}
    \rho(r) = - \frac{1}{\pi} \int_{r}^{+\infty} \frac{d \Sigma}{dR} \frac{dR}{\sqrt{R^{2}-r^{2}}},
\end{equation}
where $r$ is the 3D radius, $\rho(r)$ is the 3D density profile.
For the $\rm S\acute{e}rsic$ model of $\Sigma(R)$, the exact analytical expression of the above integral involves the Meijer G special function \citep{Mazure+2002, Baes+2011} or Fox H function \citep{Baes+2011}, both of them are complicated. Therefore, several analytical approximations are also proposed. They are described in detail in \cite{Vitral+2020} and references therein. Here we choose to use the analytical approximation given by \cite{Prugniel+1997}. In this model, one writes the dimensionless 3D density profile as
\begin{equation}
    \tilde{\rho} (x) = \frac{b_{n}^{(3-p_{\rm n})n}}{n \Gamma[(3-p_{\rm n})n]} x^{-p_{\rm n}} \exp \left[-b_{\rm n} x^{1/n} \right]
\end{equation}
where $x=r/R_{\rm e}$, $p_{\rm n}$ is a function depending on the index $n$,
\begin{equation}
    p_{\rm n} = 1-\frac{0.594}{n} + \frac{0.055}{n^{2}}.
    \label{eq:index}
\end{equation}

Therefore, in our model, the 3D luminosity density profile for the central galaxies can be written as
\begin{equation}
\begin{aligned}
    \rho_{\rm CII,g} (r) & = \frac{L_{\rm CII, g}}{4 \pi R_{\rm e, g}^{3}} \tilde{\rho} (x) \\
    & = \frac{L_{\rm CII, SFR}}{4 \pi R_{\rm e, g}^{3}} \frac{(b_n)^{(3-p_{\rm n})n}}{n \Gamma[(3-p_{\rm n})n]}  x ^{-p_{\rm n}} \exp \left[-b_n x^{1/n}  \right].
    \label{eq:galaxy 3D profile}
\end{aligned}
\end{equation}
For the extended \CII halo, since the exponential profile corresponds to the $\rm S\acute{e}rsic$ model with $n=1$, one can easily derive $\rho_{\rm CII,h}(r)$ by analogizing to Eq.~(\ref{eq:galaxy 3D profile}) and  replacing $L_{\rm CII,g}$ with $L_{\rm CII,h}$, 
\begin{equation}
\begin{aligned}
    \rho_{\rm CII,h} (r) & = \frac{L_{\rm CII,h}}{4 \pi R_{\rm e, h}^{3}} \frac{(b_1)^{(3-p_{\rm 1})}}{ \Gamma[(3-p_{\rm 1})]}  x ^{-p_{\rm 1}} \exp \left[-b_1 x  \right] \\
     & = \frac{L_{\rm CII, SFR}}{4 \pi R_{\rm e, g}^{3}} \frac{f_{\Sigma}}{ f_{R_{\rm e}}} \frac{1}{n} \frac{\Gamma(2)}{\Gamma(2n)} \frac{(b_n)^{2n} (b_1)^{(1-p_{\rm 1})}}{\Gamma[(3-p_{\rm 1})]} \\
     & \times x ^{-p_{\rm 1}} \exp \left[-b_1 x \right]. 
\end{aligned}
\end{equation}

\section{Contamination removal} \label{sec:Contamination removal}

Here we briefly introduce some foreground/interloping line removal methods for LIM in literature.  

By taking the advantage that the foreground is generally smooth in frequency space \citep{Wangxiaomin2006ApJ}, the most popular way is to fit the IR continuum of each line of sight by a polynomial, then subtract it from the line of sight spectrum to get the residual LIM signal \citep{Yue+2015}.   

Moreover, one can also separate the smooth continuum foreground and the fluctuating LIM signal through the standard principal component analysis (PCA) \citep{Yue+2015, Bigot-Sazy+2015, Cuyck+2023} or the asymmetric re-weighted penalized least-squares (arPLS) method \citep{Cuyck+2023}. 

Recently, it has been demonstrated that deep learning has the potential to be a powerful tool for removing continuum foreground, interloping lines, and noise and finally reconstructing the 3D distribution of the target emission line \citep{Moriwaki+2020, Moriwaki+2021a, Moriwaki+2021b, deep21, Zhou+2023}.

If the LIM survey area overlaps with a galaxy survey, to migrate the foreground and interloping lines, one can also simply mask the voxels occupied by bright low redshift galaxies that are suspected contamination sources \citep{Yue+2015, Silva+2015, Sun+2018, Yue+2019, Bethermin+2022, Cuyck+2023}, or blindly mask a small fraction of bright voxels \citep{Breysse+2015}. 

For removing the interloping lines from low redshift galaxies, one can also take the advantage that their cylinder power spectrum is asymmetry if they are incorrectly identified as the target emission line \citep{Lidz+2016, Cheng+2016, Yue+2019, Gong+2020}. 

\citet{Breysse2016MNRAS, Breysse+2019} also proposed that the one-point statistics can migrate the continuum foreground and the interloping lines, and get the luminosity function of the target emission line.  

\citet{Cheng+2020} proposed that, if multiple lines of sources are observed at multiple frequencies, then the target signal can be extracted by fitting the observations to a set of spectrum templates. 

Finally, the cross-correlation between \CII line and the 21 cm line can be used to distinguish the signal and contaminations \citep{Gong+2011, Silva+2015}, while the cross-correlation between foreground lines and galaxy surveys are also explored in \citet{Silva+2015} in order to probe the intensity of the foregrounds.

%%%%%%%%%%%%%%%%%%%%%%%%%%%%%%%%%%%%%%%%%%%%%%%%%%

% Don't change these lines
\bsp	% typesetting comment
\label{lastpage}
\end{document}